\documentclass[conference]{IEEEtran}
\usepackage{amsfonts}

\usepackage{amsmath}
\usepackage{centernot}

\usepackage{hyperref}

\ifCLASSINFOpdf
\else
\fi
\usepackage{mathtools}

\usepackage{graphicx}

 \usepackage{booktabs}

\usepackage{listings}

\usepackage{multirow}

\begin{document}

\title{Toward Refactoring of DMARF and GIPSY Case Studies a Team 7 SOEN6471-S14 Project
Report}


\author{\IEEEauthorblockN{Abdulrhman Albeladi}
\IEEEauthorblockA{Concordia University\\
Montreal, Canada\\
blady911@gmail.com}

\and
\IEEEauthorblockN{Ahmed Almessabi}
\IEEEauthorblockA{Concordia University\\
Montreal, Canada\\
salmessabi@gmail.com}

\and
\IEEEauthorblockN{Aber Abozkhar}
\IEEEauthorblockA{Concordia University\\
Montreal, Canada\\
aberabozkhar2@gmail.com}

\and
\IEEEauthorblockN{Huda Mohamed}
\IEEEauthorblockA{Concordia University\\
Montreal, Canada\\
 h.m.haddar@gmail.com}
\and
\IEEEauthorblockN{Jilson Thomas}
\IEEEauthorblockA{Concordia University\\
Montreal, Canada\\
jillztom@gmail.com}
\and
\IEEEauthorblockN{Zakaria Alomari}
\IEEEauthorblockA{Concordia University\\
Montreal, Canada\\
zakaria.alomari@gmail.com}
 }


%Telephone: (800) 555--1212, Fax: (888) 555--1212}


\maketitle

\begin{abstract}
Software architecture is defined as the process of a well-structured solution that meets all of the technical and operational requirements, as well as improving the quality attributes of the system such as  readability, Reliability, maintainability, and performance. It involves a series of design decisions that can have a considerable impact on the system’s quality attributes, and on the overall success of the application. In this work, we start with analysis and investigation of two open source software (OSS) platforms DMARF and GIPSIY, predominantly implemented in Java. Many research papers have been studied in order to gain more insights and clear background about their architectures, enhancement, evolution, challenges, and features. Subsequently, we extract and find their needs, high-level requirements, and architectural structures which lead to important design decisions and thus influence their quality attributes. Primarily, we reversed engineering each system$'$s source code to reconstruct its domain model and class diagram model. We tried to achieve the traceability between requirements and other design artifacts to be consistent. Additionally, we conducted both manual and automated refactoring techniques to get rid of some existing code smells to end up with more readable and understandable code without affecting its  observable behavior.

\end{abstract}

%\begin{IEEEkeywords}
%Computer Society, IEEEtran, journal, \LaTeX, paper, template.
%\end{IEEEkeywords}

% \fi
%\IEEEpeerreviewmaketitle

%\section{Introduction}
%This demo file is intended to serve as a ``starter file''
%for IEEE conference papers produced under \LaTeX\ using
%IEEEtran.cls version 1.8 and later.
%% You must have at least 2 lines in the paragraph with the drop letter
%% (should never be an issue)

%Give a short description of your study. More importantly, describe the motivation for your study.
%I wish you the best of success.

\section{Introduction}

The main goal of this work is to get better understanding and deep comprehension of the architecture of two case studies, Distributed Modular Audio Recognition Framework DMARF and General Intentional Programming System GIPSY, implemented in JAVA. To achieve this goal, we begin with studying different papers to analyse and investigate these two open source software (OSS), and to gain more insights and clear background about their architectures. We start by extracting and summarizing the core frameworks’ design artifacts for both OSSs such as; high-level requirements, fully dressed use cases, domain diagrams, and class diagrams. Next, we perform another step of identification of code smells.  Some of these code smells are identified manually; others are identified automatically with the help of some tools such McCabe IQ, Logiscope, and JDeodorant.  Subsequently, we apply different refactoring techniques on the existing code smells in order to get rid of these smells, and to restructure the code without changing its external behaviour, thus increase the readability, understandability and reduce the complexity to make the code more maintainable as well as extensible. In addition, we come across implemented design patterns in both case studies, DMARF and GIPSY that deal with a specific problem in the design or implementation of software systems. Patterns can help us to include the existing well proven coding methods in software development which follow a good design methodology. Some of these patterns are Factory, Observer, Singleton, Adapter, Facade and many others. Lastly, we conduct JUnit test-cases to ensure that the applied refactoring techniques don’t change the system’s external behavior.

The paper is structured as follows: Section 1 presents two case studies DMARF and GIPSY predominantly written in Java with the goal of finding their needs, high-level requirements,  and architectural structures.  Also, it provides initial estimations of the size of both case studies. Section 2 defines the high level requirements, fully dressed use cases, and conceptual domain models. In addition, It shows how both systems can be merged and fused each other, where DMARF could use GIPSY$'$s run-time for distributed computing instead of its communication technology. Section 3 presents UML class diagram for each system and the relationships between the classes in each diagram. Additionally, it shows how the traceability concept is achieved between requirements and other design artifacts to end up with a consistent design. List of  the identified code smells and specific refactoring techniques will be used to include it in that section. Section 4 presents the implementation of refactoring and test cases as well as the conclusions of some insights gained from our experiments.

\section{Background}
\subsection{OSS Case Studies}
%----------------------------------------------
\subsubsection{DMARF}
Distributed Modular Audio Recognition Framework (DMARF) is a distributed version of the original Modular Audio Recognition Framework (MARF). In order to illustrate DMARF, MARF first needs to be illustrated. As in [1] MARF is a Java open-source project for patterns recognition, signal processing, and natural language processing (NLP) [1]. MARF can run stand-alone, over the network, or used as an application$'$s library.  There are several MARF$'$s applications, like SpeakerIdentApp and FileTypeIdentApp [2]. MARF$'$s architecture is a sequential pipeline with lake or even no concurrency when having a task of processing a bulk of voice samples. This problem has been resolved by extending the classical MARF to DMARF [3].
 
The pipeline stages or algorithms are designed in a modular way and provide extensibility feature to allow adding more algorithms. MARF contains pipeline stages that communicate with each other in order to obtain the required data. As shown in Figure 1 the pipeline consists of the four basic stages: sample loading, preprocessing, feature extraction, and training or classification [1]. In sample loading stage, the sample is first loaded and then converted to a supported type, for instance WAVE. Then, this sample is sent to preprocessing stage, where the sample gets normalized and filtered accordingly, so it can be prepared for feature extraction, which will be done in the next stage. After feature extraction comes training and classification, where the feature vectors could be considered as training data been stored or classified and the system learn from it [2].

The goal of extending classical MARF to DMARF as presented in [3] is to distribute  the pipeline and make it runnable over a different groups of loosely coupled computers that work together closely. This flexibility of distribution these services aims to offload the bulk of multimedia or data to be processed to a higher performance servers that able to communicate while the collection of data can be done at several low-cost computers or PDAs without need to have processing and storage capacity, just they pass the collected data to the servers [3]. 

Yet another enhancement added to DMARF has been introduced in [4] by  implementing disaster recovery, replication techniques and communication technology independence, such as RMI, CORBA, and Web Services (WS), which will allow the pipeline stages to communicate. WS makes the components even more interoperable and platform-independent, also, WS implementation in DMARF makes it even more widely available over the Internet.

DMARF is divided into layers, front-ends and back-ends application services, where all MARF$'$s pipeline stages are placed in the front-ends, and the disaster recovery and replication techniques are implemented in the back-ends as shown in Figure 2. The front-end exists on the client side and on the server side. On the client side, a client application invokes services from the front-end on the server side, which means connect and query the servers. When a client application invokes services from the front-ends, the front-ends$'$ services invoke neighbor services or the back-ends$'$ services [2]. On the server side, the front-end services invoke other services from the back-end. At the same moment, the services like disaster recovery and service replication are a back-end for the client [4]. 
 
To achieve the management over MARF services [3], Simple Network Management Protocol SNMP is used to be integrated with use of common network devices and service to provide the administrators with the capability to manage MARF nodes by using a familiar protocol. Moreover, it monitors and controls their performance, gathers statistics, and sets desired configuration. In contrast, DMARF’ components are stand-alone and distributed among different computers, and communicate to each other via RMI, XML-RPC, CORBA, and TCP connections, which leads to not understanding of SNMP Protocol.  So each managed service in DMARF has to have: 

\begin{itemize}
\item A proxy SNMP-aware agent for management tasks. 
\item A delegate instrumentation proxy to communicate with the specified service.

\end{itemize}

%zakerai    ______----------*******
There are different points of view related to Distributed MARF: First, applying Autonomic Computing makes a system is self-managing autonomic system so that ongoing software and hardware complexity is decreased.  Two properties of self-managing: self-healing and self-optimization are applied in DMARF, self-protecting autonomic property is also needed [1].  

Second, a self-forensics considers as an autonomic property to boost the Autonomic System Specification Language (ASSL) framework of formal specification tools for autonomic systems to add the self-forensics autonomic property to enable generation of the Java-based Object-Oriented Intentional Programming language code linked with traces of Forensic Lucid for encoding contextual forensic evidence and other expressions. ASSL formal modeling, specification, and model checking has been applied to a number open-source, academic, and research software system specifications such as the Distributed Modular Audio Recognition Framework (DMARF) [5].

Third, Security can be breached by vicious attacks on the distributed systems such as DMARF, so some of the vicious attacks can be tackled by using middleware technologies while the rest can only be addressed by implementing robust security system. Thereby, JDSF security framework has been used as a security layer once communicating with nodes external to the local area network. JDFS resolves issues related to confidentiality, integrity, authentication, and availability. The proposed results of JDSF do blanket a wide cluster of the goals leaving malicious code detection unsolved [6].

The domain of DMARF system which is basically  Audio and Voice Recognition Framework for patterns recognition, signal processing, and natural language processing (NLP) [2], [5]. Furthermore, DMARF extends to be used in different domains other than audio and voice processing such as speech recognition, forensics, security applications, text-independent Speaker-identification, language identification, natural language probabilistic parsing, and other classification applications. It also acts as a library in different applications or used as a source for learning and extension. Most of these applications are revolved around  the  multifaceted approach provided by DMARF.  For example, one of DMARF’s applications is SpeakerIdentAppt that  has a database of speakers, where it can identify who people are regardless what they say. This application will extremely useful  in law enforcement agencies and police department for forensic analysis that need to identify speakers across all jurisdiction [3], [4].

 \begin{figure} [ht!]
  
  \centering
    \includegraphics[width=0.5\textwidth]{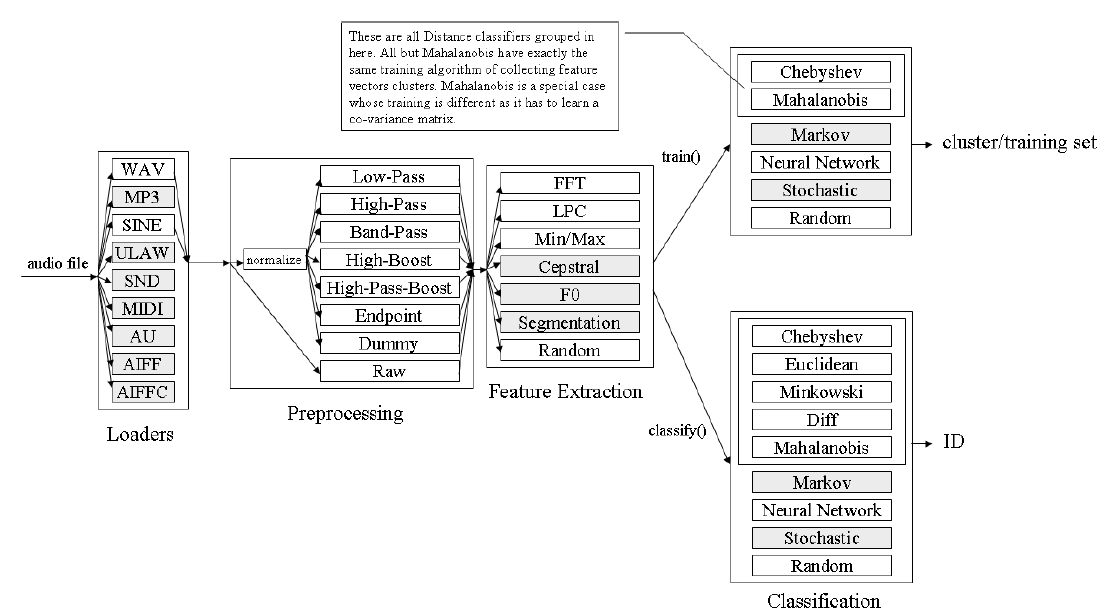}
\caption{MARF$'$s Architecture [3, p.3] }
\end{figure}

 \begin{figure} [ht!]
  
  \centering
    \includegraphics[width=0.5\textwidth]{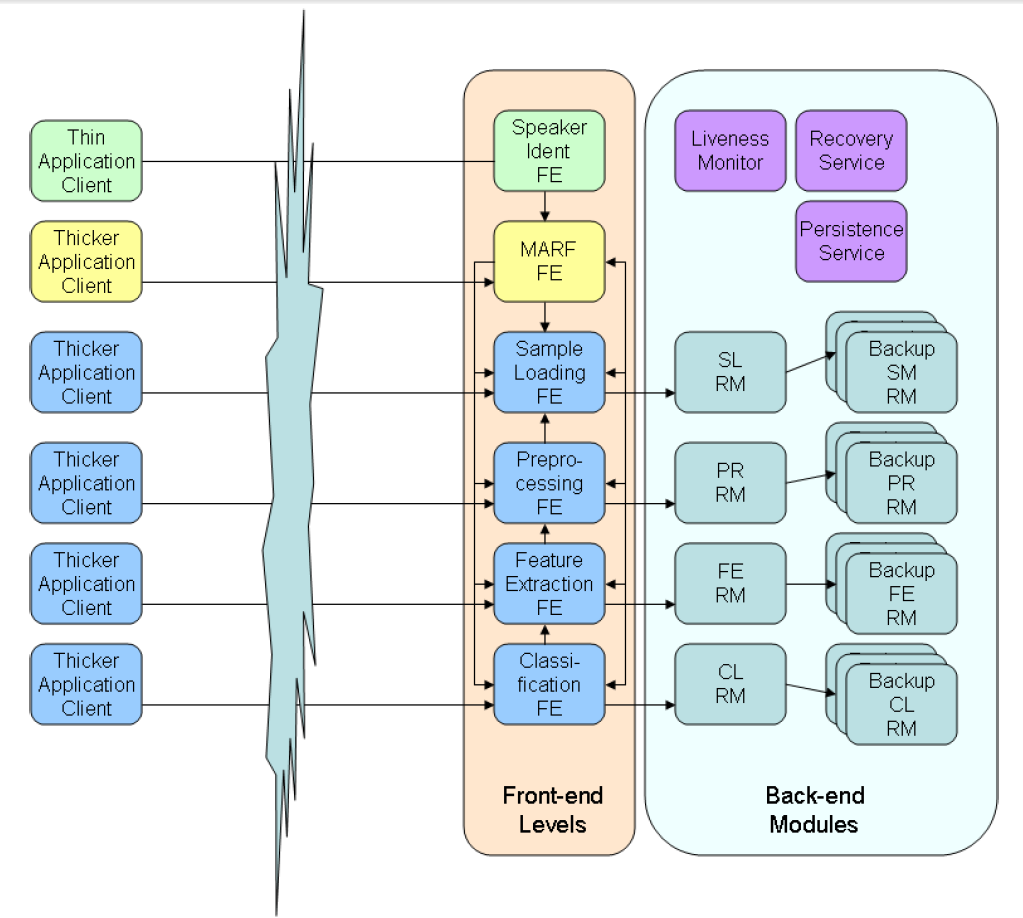}
\caption{DMARF$'$s Architecture [3, p.4]}
\end{figure}

%-----------------------------------------------------------

\subsubsection{GIPSY}
General Intentional Programming System (GIPSY) is a distributed demand-driven evaluation engine. The purpose of implementing GIPSY is to have a flexible compiler architecture that could read and compile multiple intentional programming languages [11]. Although the evolution of Intentional Programing IP has achieved a great maturity, there was a limitation in previous supported tools for this type of programming as well as lack of IP$'$s visibility lead to the implementation of GIPSY system [9].The main goal of GIPSY is to adapt to the fast development and the great diversity of the intentional family of programming languages. Furthermore, through hybrid programming in combination with standard procedural languages, it allows reusing of legacy code, and enables the use of different distributed execution middleware at run-time. The system provides these characteristics easily, in a way that makes the users can create new Intentional Programming Languages, use of different procedural languages, or middleware technologies [8].

 GIPSY$'$s architecture is a multi-tier architecture and have the advantage of using modular development for the compiler components. It is composed of three main components, namely General Intensional Programming Compiler (GIPC), General Education Engine (GEE) and Intensional Run-time Programming Environment (RIPE). Each component has its own architecture or design which means independent and can be replaced or maintained without affecting the other components. In addition, the subsystems are designed towards generality, flexibility and efficiency [9]. 

The GIPC component, is the main compiler for GIPSY, and in charge of translating IPs Program to C, which in turn will be compiled in a standard way. The GEE’s architecture is a distributed multi-tier architecture, which allows one or more instances for each tier. This architecture is similar to peer-to-peer architecture, which makes GIPSY more robust when a failure occurs in any trier or node. GEE component is mainly concern about generating tasks that can be executed in a parallel way using the computation model of demand-driven model. The GEER is language independent and a run-time resources dictionary compiled from GIPL  program. GEER’s instances are created by GIPC when it compiles a program [11], [7]. Figure 3, Shows GIPSY Architecture.

 \begin{figure} [ht!]
  
  \centering
    \includegraphics[width=0.5\textwidth]{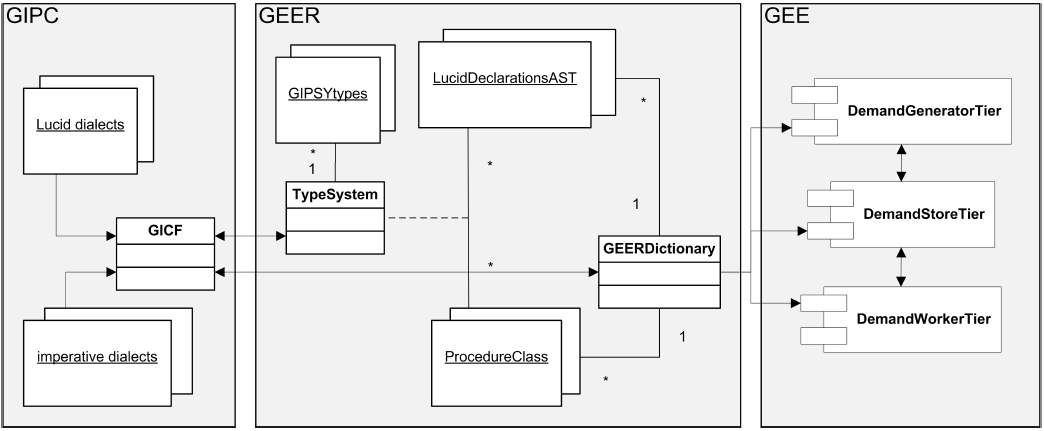}
\caption{The GIPSY Architecture [2, p. 148]}
\end{figure}
 
GIPSY has shown an effectiveness in meeting its main purposes mentioned above, as well as it is evolved in a flexible way. An example of that, GIPSY provide context-oriented multidimensional reasoning of intensional expressions without limiting the scope of evaluation of math expressions as GIPSY is based on the higher-order intensional logic (HOIL). HOIL combines functional programming with different intensional logics to allow explicit context expressions to be evaluated in GIPSY [11].

GIPSY solves the distributed architecture problems of the generic runtime system. Previously, the architecture was not fully integrated and detailed working flow had to be modified [10]. The proposed solution for the above mentioned problems are based on the GLU’s generator-worker architecture. Through multiple design iterations, it extended to be a multi-tier architecture and applied the high level designs using wrapper classes for each newly introduced tier type, mainly DWT (Demand Worker Tier), DGT(Demand Generator Tier), DST (Demand Store Tier), and GMT(General Manager Tier). This Solution is an evolution of the GIPSY runtime system’s original architecture.

Another aspect that GIPSY consider is investigating hybrid programming language in local and distributive way. Therefore, a Demand Migration Framework (DMF) is designed and introduced in the execution environment of GIPSY in order to provide a generic and dynamic infrastructure for the purpose of distributed demand-driven communication [7]. DMF was built using two different java middleware technologies, Jini (Java-based and service-oriented middleware technology) and JMS (Java-based and message-oriented middleware technology). However, the two middlewares are separated from GIPSY and from each other. In order to make the GEE scalable enough to compute the heavy hybrid programs in a distributive way, there was a necessity to unify the two java middleware technologies, Jini and JMS. This has been resolved by doing refactoring. First, by making JMS and Jini part of the same interchangeable framework$'$s implementation. Redefine the role of demand dispatcher and Transport Agent and Compare Jini and JMS with JVM, compare their APIs and compare the ease of deployment and startup [7].

Based on the conceptual level, implementing the self-forensics within the Autonomic System Specification Language (ASSL) and General Intensional Programming System (GIPSY) is described through their founding core works. ASSL formal modeling, specification, and model checking has been applied to a number of open-source, academic, and research software system specifications, such as GIPSY. ASSL is a framework that provides a multi-tier specification model for Autonomic System (AS) as well as ASSL allows expressing ASs, as a set of interacting Autonomic Elements (AEs), at three main levels: AS level, AS Interaction Protocol (ASIP) level, and AE level [12].

GIPSY has illustrated  the effectiveness of  intentional programming to solve critical problems in various domains  such as tensor programming, distributed operating systems, and software versioning, hybrid programming language, demand driven eductive, and forensic.

\subsubsection{Summary}

MARF services uses the implementation of three components such as RMI, COBRA and Web services (WS). But, RMI and COBRA has got some problems such as  need of RemoteException throws during stubs creation and generation of data structures for COBRA. For a greater portability using HTTP service for the MARF, this WS is implemented for the DMARF’s COBRA and RMI. This WS can even communicate with these two DMARF components even through plain UDP or TCP protocols. An algorithm with ASSL of the pipeline pattern recognition of the DMARF is used to create a ADMARF self-protection specification model which will be very much functional in autonomous environments. However, it is not a complete autonomous ADMARF specification model. For the minimal testing, generation of proxy agent and compilation is enough in debug form of MIBs. To allow self-forensics property implementations in the ASSL toolset, some preliminary work is done. To protect the security aspects of many uncontrolled public networks, a solution of JDSF is proposed and it covers a wide range of security aspects for scientific research and experimental distribution systems. 

%******GIPSY conclusion
The multi-tier components will offer a high scalability and flexibility to the GIPSY at runtime when it is implemented and fully tested. Some new packages were developed and defined for this multi-tier implementation. The design of GIPSY architecture allows components to be replaced at run-time or at compiler-time so as to improve flexibility and scalability and allows automated generation of hot spots for support of programing language. In the area of flexible hybrid intentional research programing, the GIPSY is well off for the scientists and researchers. In the GIPSY programming paradigms, the central value is the concept of context as a first class which explores the lucid programing language family. The distributed transport of the demands’ DMS has implementations in Jini, plain RMI, and JMS and the POC integration of the middleware is implemented using Jini and JMS. In the future, the other components in the GIPSY can be improved to work with unison by refactoring and cleaning up the code.

The methodology used to analyze both DMARF and GIPSY projects source code was simple by using a tool called SonarQube [13]. SonarQube (previously known as Sonar which mean is the central place to manage code quality, offering visual reporting on and across projects and enabling to replay the past to follow metrics evolution [14]) is an open web-based application platform to manage code quality. We used SonarQube 4.3.2 Released which fixes migration issues that can occur on specific cases like Migrations that convert technical debt from hours to minutes are too slow[15]. Also we used Sonar Runner v2.4 This version provides various improvements and bug fixes[15].

Sonar Server has been used in order to interpret and analyze both projects considering Apache HTTP Server to be associated with Sonar Server.  Afterwards, each project has used sonar-project-properties file to set sonar properties. Thereby, the team has corporated to run and access Sonar Runner by command line as well as Sonar Runner reads the selected project files and collected required measurements. As a result, the values is presented in a browser. In order to get theses measurements, Sonar did not compile niether systems. Also, the measurements were only on Java files. 

 The required measurements are presented in Table \ref{tab:1} as follows:

 \begin{table}[htpb]
    
  \renewcommand{\arraystretch}{1.3}
 \caption{ DMARF And GIPSY Measurements} 

    \begin{tabular}{|c|c|c|}
%{|p{3cm} | p{2cm} | p{2cm}|}

    \hline
    Measurements & DMARF & GIPSY\\ \hline
   Files  &  948   &  608 \\ \hline

    Packages  & 377& 164       
 \\ \hline
 
  Classes  & 978& 666       
 \\ \hline

  Methods  & 6760& 6276       
 \\ \hline
  Line of Text  & 123261& 140828       
 \\ \hline

  Number of statements  & 28478& 56078       
 \\ \hline
  Comment Lines  & 19939 & 13814       
 \\ \hline
    Line of code  & 73393 & 104199       
 \\ \hline

    \end{tabular}

\label{tab:1}

 \end{table}

Table \ref{tab:1} , it demonstrates several measurements for two different open-source systems DMARF, and GISPY respectively. After executing SonarQube, it shows DMARF system exceeds slightly less the half of files and classes number than GIPSY. Likewise, DMARF in number of packages is greater by three times than GIPSY system. In terms of number of methods, both systems seem quite close to each other considering DMARF is little bit bigger whereas line of text in GIPSY considers less than DMARF. Moreover, number of statements and comments lines appear away bigger in DMARF than the other whereas line of code of DMARF is less than GIPSY. Thus we conclude that DMARF is more complex as compared to GIPSY because DMARF’s number of files, classes and methods are higher.

\section{Requirements and Design Specifications}
\subsection{Personas, Actors, and Stakeholders}

\subsubsection{DMARF}

\mbox{} \\
\noindent
{\bf Persona:}

Ahmad Aljohani is 35 years old.  Ahmad is a male adult, and have a good experience with computers and the Internet. He Received Masters degree in Software Engineering from Concordia University.  Ahmad now is working as a developer at CIT company, and have attended many workshops and conferences in how to develop well designed software systems. Also, he has a good  knowledge in developing and integrating software systems, and has develop couple software systems. He is currently interested in developing a survey software system, that would allow its users to create surveys and distribute them. This software will have an option to allow its users to get authenticated through voice. This kind of feature would needs to identify its users through voice recognition and get them authenticated when there is a match. He does not have enough time to implement such a feature, instead he needs a good library system that can identify users via voice, and match them to the existing users to get them authenticated. In addition, he needs that library to be simple to use and get benefit from it. He never liked to use a library that does not have a good documentation or user manual. In addition, It would create a big problem for Ahmad if the system did not identify accurately the right users for the right accounts. Performance is an essential quality for Ahmad$'$s software system because if it is not fast enough, his users would not use his system. Ahmad get frustrated easlly when he uses complex systems$'$ interface. Moreover, he never liked to ask his colleagues to teach him how to use a system because he thinks it would make him look stupid. Ahmad wants his system to be portable, so it can run in different platforms.
\mbox{} \\

\noindent
{\bf Actors:}

Writer Identification Application processes tasks of scanned hand-written documents to identify the writer, such as students$'$ exams verification and personal checks identification. This application can achieve its goal by using MARF$'$s approach to define a common set of integrated APIs for the pattern recognition pipeline [16]. This technique is useful for biometric modality with applications in forensic and historic document analysis [17]. Writer Identification Application is considered as a primary actor for MARF.

JDSF allows to work with several types of data storage in order to anticipate different level of security algothrims and methodologies in certain environment.  JDSF covers several aspects to keep data privacy, to integrate and to authenticate data stemming from a trusted source for DMARF. Once an attack takes place somehow, JDSF shall protect DMARF$'$s data and secured them. Besides, It does capture the configuration data related to setting connection or any other properties related to the distributed systems [6]. Therefore, JDSF  is considered as a secondary actor.

Text-Independed Speaker Identification Application processes amount of voice samples to test MARF$'$s functionalities so this application tells who, gender, accent, and spoken language of the speaker using MARF$'$s pipeline.  Thereby, Text-Independed Speaker Identification application contributes to share services, to arrange as a modular, and to facilitate adding new algorithms for use or experiments. JSDF uses these algorithms implemented to recognize patterns and to process tasks. Not only does that, but also it is leveraged from MARF$'$s backbone particularly pipeline to get data needed [18]. Thus we considered this application as a primary actor. 

Communication technology is a standard architecture for a distributed system, and it is designed to allow distributed systems like GISPY and DMARF to interoperate in a heterogenous environment, where systems can be implemented in different programming languages and/or deployed on different platforms. It is based around  the concept of a client application using the services available on a remote machine, or server. The object$'$s interface represents a contract between the client and the server. This interface is written as a Java  interface for Java RMI, in IDL for CORBA, and in WSDL for web services [21]. Therefore, communication technology can be considered as a secondary actor for DMARF and GIPSY.

\mbox{} \\

\subsubsection{GIPSY}

\mbox{} \\
{\bf Persona:}

Anton is a 34 year old Manipal University professor and a freelance software developer. He has a Phd in the field of software engineering and has many experience in software versioning and hybrid, tensor programmings. He has got an industrial experience of 7 years as the lead/senior developer for a famous software developing company and very good knowledge in Java, C, C++ and a limited knowledge in .NET. Currently he is trying to develope a fingerprint recognition application for the Police Department. He wants a platform which is independent of the programing language to compile and execution since he might use different languages to develop the system. Therefore, there should be a flexible compiler for the system. It should also be compatible to solve any security issues that may arise in the fingerprint recognition environment. Also, he is experiencing some problems with the distributed architecture systems that he is working on as part of the project. Currently, he is looking for a system that could encode an image sample into a Forensic lucid language, and then get compiled by GIPSY to investigate the fingerprint. The developing system should be made up of different components which will further interact to each other. Therefore, he is mainly interested in a modular development for the system components. 

 \mbox{} \\
\noindent
{\bf Actors:}

Software Versioning system merge all versions that have a similar version description into a single version, such as Lemur system using an intensional versioning technique. Lemur system uses intensional programming implemented in a demand-driven computation framework. It can achieve its goal by using GIPSY techniques [9], [11]. Software Versioning system is considered as a primary actor for GIPSY.

Hybrid programming: is an integration of intensional and imperative languages that is required for different needs such as reusing legacy code. GIPSY provides developers with a platform where they can create new Intensional Programming Language IPL, enables the use of different procedural languages, or middleware technologies in an easy way with minimal technical knowledge by increasing the automation of the extension process [19]. The developers are considered as primary actor.

Finger Print Identification App is an application that recieves an image sample from the target system, like in our situation Crime Investigation system. Then, it encodes that image into Forensic Lucid language. Later, GIPSY recieves the Forensic Lucid code. Then, GIPSY compiles and evaluates it.The Finger Print Identification App is considered as primary actor.

\subsection{Use Cases}

\subsubsection{DMARF}
\mbox{} \\

\noindent
{\bf Use case name:} Identify Speaker$'$s Voice 

\noindent
{\bf Scope:} System under investigation.

\noindent 
{\bf Level:} User level

\noindent
{\bf Primary Actor:} Survey System.

\noindent
{\bf Stakeholders and Interests:} Survey System:

\begin{itemize}
\item Wants the {\bf SpeakerIdentApp} to \underline{identify} the user accurately.
\end{itemize}

\noindent
{\bf Preconditions:}

\begin{enumerate}
\item {\bf Survey system} has a {\bf voice sample} ready to be uploaded.

\item Both {\bf Survey system} and {\bf SpeakerIdentApp} are connected to the Internet. 

\end{enumerate}

\noindent
{\bf Postconditions:}
\begin{enumerate}
\item  A {\bf user} gets identified.

\item  A reply with the matching result.

\end{enumerate}

\noindent
{\bf Main success scenario:}\\
\vspace*{-0.2in}

\begin{enumerate}

  \item  {\bf Survey system} wants to \underline{identify} a {\bf user}.

  \item  {\bf SpeakerIdentApp} asks to \underline{upload} the {\bf voice sample}.

  \item  {\bf Survey system} \underline{upload} the {\bf voice sample}.

\item  {\bf SpeakerIdentApp} \underline{processes} the {\bf voice sample}.

\item  {\bf SpeakerIdentApp} \underline{matches} it with registerd users in {\bf Survey system}.

\item  {\bf SpeakerIdentApp }\underline{replies} to {\bf Survey system} with the matching result.

\end{enumerate}

\noindent
{\bf Extensions:}

\noindent
The system shall crash at any time, the user needs to close and reopen it.

\noindent
1a: Network connection fail: 
      \mbox{} \\

 1a.1: before uploading voice file:
               
 \begin{enumerate}
\item system user should looks for good enough connection.

\item Go to a step 1.
\end{enumerate}
      \mbox{} \\
1a.2 After uploading voice file: 
         
 \begin{enumerate}
\item Go to step 2
\end{enumerate}
      \mbox{} \\
3a: Survey system fails to upload the sample.
      
 \begin{enumerate}
\item  SpeakerIdentApp asks to upload the voice sample again.
\item  Go to step 3.

\end{enumerate}
      \mbox{} \\
4a: SpeakerIdentApp fails to process the voice sample due to Network Failure:
       
 \begin{enumerate}
\item SpeakerIdentApp looks again for a new or former connection.
\item  Go to step 2.
\end{enumerate}
          \mbox{} \\
5a: SpeakerIdentApp fails to match the voice sample with any registered users.
  
\begin{enumerate}
\item  SpeakerIdentApp reply with an error message to survey system that it failed to identify any user.

\item  SpeakerIdentApp asks to upload the voice sample again.

\item  Go to step 3.
\end{enumerate}
      \mbox{} \\

\noindent
{\bf Special requirements:}
\begin{itemize}
\item Voice sample must be clear.
\item  Voice sample must be uplaoded within 15 second. 
\item Confirmation of SpeakerIdentApp (or reason for failure) to be provided to the Survey system within 20 seconds of submission. 
\end{itemize}
      \mbox{} \\

\noindent
{\bf Technology and data variations list:}
\begin{itemize}
\item The Survey system should be able to upload arbitrary type of voice sample. 
\end{itemize}
      \mbox{} \\

\noindent
{\bf Frequency of occurrence:}
\begin{itemize}
\item Could be nearly continuous. 
\end{itemize}
      \mbox{} \\

\noindent
{\bf Miscellaneous:}
\begin{itemize}
\item  What if the Survey system could not be able to upload the voice sample? 
\item  What if the SpeakerIdentApp did not identify the right user for the right account? 
\item  What if SpeakerIdentApp does not support all list of voice extensions?
\end{itemize}
      \mbox{} \\

\subsubsection{GIPSY}
\mbox{} \\

\noindent
{\bf Use case name:} Identify Fingerprint  

\noindent
{\bf Scope:} System under investigation.

\noindent 
{\bf Level:} User level

\noindent
{\bf Primary Actor:} {\bf Crime Investigation System}

\noindent
{\bf Stakeholders and Interests:} Crime Investigation System: 
\begin{itemize}
\item Wants to \underline{identify} the fingerprint image accurately.   
\end{itemize}

\noindent
{\bf Preconditions:}

\begin{enumerate}
\item  {\bf Fingerprint image} is {\bf scanned} and ready to upload.
\item Both {\bf Crime Investigation system} and {\bf Fingerprint Identification system} are connected to the Internet.

\end{enumerate}

\noindent
{\bf Postconditions:}
\begin{enumerate}
\item  A suspect has been identified from his/her fingerprint.

\item  A result of matching or not has been sent to the police system.

\end{enumerate}

\noindent
{\bf Main success scenario:}\\
\vspace*{-0.2in}

\begin{enumerate}

 \item  The {\bf Crime Investigation system} wants to \underline{identify} a suspect’s fingerprint.

 \item  The {\bf Fingerprint Identification system} asks to \underline{upload} the {\bf image}.

  \item  {\bf Crime Investigation system} \underline{uploads} the {\bf image} to the {\bf Fingerprint Identification system}. 

\item  {\bf Fingerprint Identification system} \underline{processes} the {\bf image}.

\item  {\bf Fingerprint Identification system} \underline{compares} the uploaded {\bf image} with the registered suspects.

\item  {\bf Fingerprint Identification system} \underline{replies} with the matching result.
\end{enumerate}

\noindent
{\bf Extensions:}
 
\noindent
3a. The Crime Investigation system failed to upload the image.
 \begin{enumerate}
\item   Fingerprint Identification system displays an error message.
\item  Fingerprint Identification system asks to upload the image again.
\item  Go to step 3.
\end{enumerate}
      \mbox{} \\

\noindent
  4a. If the Fingerprint Identification system fails to process the image because of insufficient network connection.
 \begin{enumerate}
\item  Fingerprint Identification system finds an appropriate network connection. 
\item  Go to step 3.
\end{enumerate}
      \mbox{} \\
\noindent
 5a. If the Fingerprint Identification system does not recognize the fingerprint image.
    	 
\begin{enumerate}
\item    Fingerprint Identification system responds with a message that no suspect was found.
\item  Go to step 2.
\end{enumerate}
      \mbox{} \\

\noindent
{\bf Special requirements:}

\begin{itemize}
\item The image should be clear enough to be processed
\item  Sufficient time for upload the image should not be more than 20 seconds. 
\item The system should be secure enough to prevent any incorrect usage.
 \item  Confirmation response should not exceed more than 20 seconds.
\item    Robust recovery for the system in case of failure
\end{itemize}
      \mbox{} \\

\noindent
{\bf Technology and data variations list:}
\begin{itemize}
\item The system allows various image format readings
\end{itemize}
      \mbox{} \\

\noindent
{\bf Frequency of occurrence:}
\begin{itemize}
\item Might occur frequently. 
\end{itemize}
      \mbox{} \\

\noindent
{\bf Miscellaneous:}
\begin{itemize}
\item  What if a problem happened while uploading a fingerprint image?
\item  What if the image format is not supported by the Fingerprint Identification system? 
\end{itemize}
      \mbox{} \\

\subsection{Domain Model UML Diagrams}

\subsubsection{DMARF}
\begin{figure*} [ht!]
  \centering
    \includegraphics[width=0.7\textwidth]{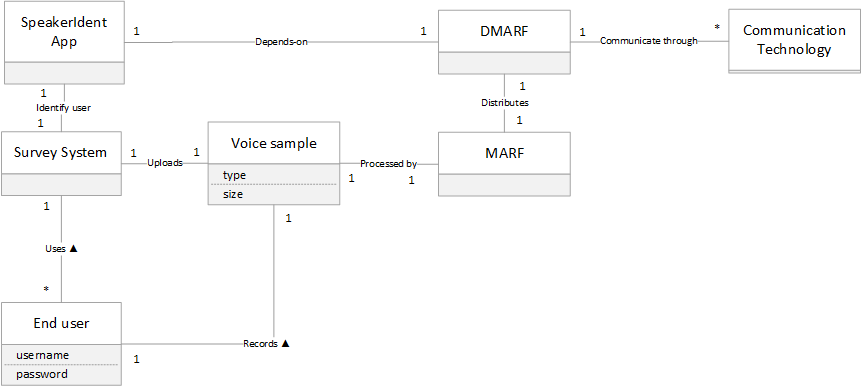}
\caption{DMARF$'$s Domain Model}
 \label{fig:D1}
\end{figure*}

Figure ~\ref{fig:D1} , Shows the DMARF$'$s domain model with {\bf SpeakerIdentApp}. The {\bf Survey system} can be used for any type of survey related applications. It needs the {\bf end users} to get authenticated by their voices to use the system. A {\bf SpeakerIdentApp} is used for identifying the end users from their recorded {\bf voice samples}. This identification will be done by using the sound recognition in {\bf MARF}. First, the users will be asked to record their voice on login process in the {\bf Survey system}. This {\bf voice sample} will be recorded by the end user through the {\bf Survey system} and uploads to {\bf DMARF} through the {\bf SpeakerIdentApp}. It is then moved to {\bf MARF} to compare it with the already existing voice records.

The {\bf MARF} pipeline is used to compare and to process the pattern recognition and to send the feedback back to the {\bf SpeakerIdentApp} and then to the {\bf Survey system}. If the {\bf MARF} gives a positive reply back to the {\bf SpeakerIdentApp}, and then to the {\bf Survey system}, the user will be authenticated and be given permission to access the {\bf Survey system}. If the {\bf MARF} gives a negative feedback or if the patterns do not match, then the {\bf Survey system} will not authenticate that a particular user from accessing the system. Only one voice sample can be given to the {\bf Survey system} at a time by one user. The {\bf Survey system} can be used by many users at the same time.

\begin{figure*} [ht!]
  
  \centering
    \includegraphics[width=0.7\textwidth]{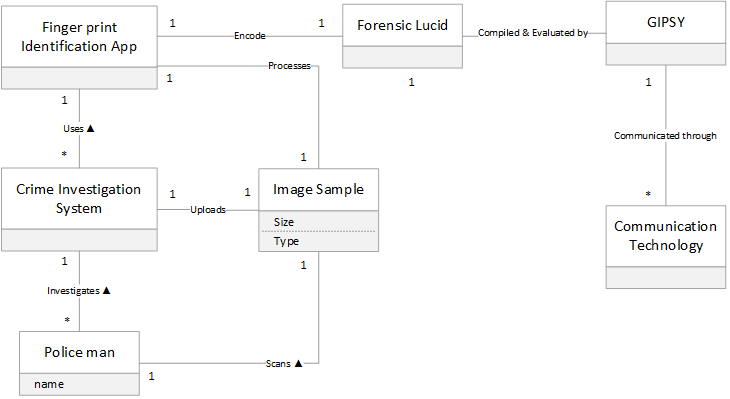}
\caption{GIPSY$'$s Domain Model}
 \label{fig:D2}
\end{figure*}

\subsubsection{GIPSY}
Figure ~\ref{fig:D2} , Shows the GIPSY$'$s domain model. Using a {\bf Fingerprint Identification application} a {\bf crime investigation system} can identify the {\bf fingerprints} of a suspicious person who already has a record. The end users will be {\bf the police officers} who are investigating any crimes. Many {\bf policemen} can use the {\bf crime investigation system} to scan and upload different {\bf fingerprints} at the same time to check a suspicious {\bf fingerprint}. {\bf Crime Investigation System} will demand {\bf Fingerprint Identification application} to identify the recent image. The {\bf Fingerprint Identification application} will assign an image of {\bf fingerprint} to {\bf Forensic Lucid expression}, which already has been encoded the image (evidence) in the consistent syntax. In {\bf GIPSY}, GIPC will be responsible for parsing such {\bf Forensic Lucid specification} and GEE will be responsible for executing it. By executing the {\bf Forensic Lucid}, it will match the encoded evidences/images with the new encoded image of the suspicious {\bf fingerprint}. An evaluation of whether it matched or not will be sent to the {\bf Police officer}. {\bf GIPSY} requires some {\bf communication technology} services such as RMI, WS for communication.

\begin{figure*} [ht!]
  
  \centering
    \includegraphics[width=0.7\textwidth]{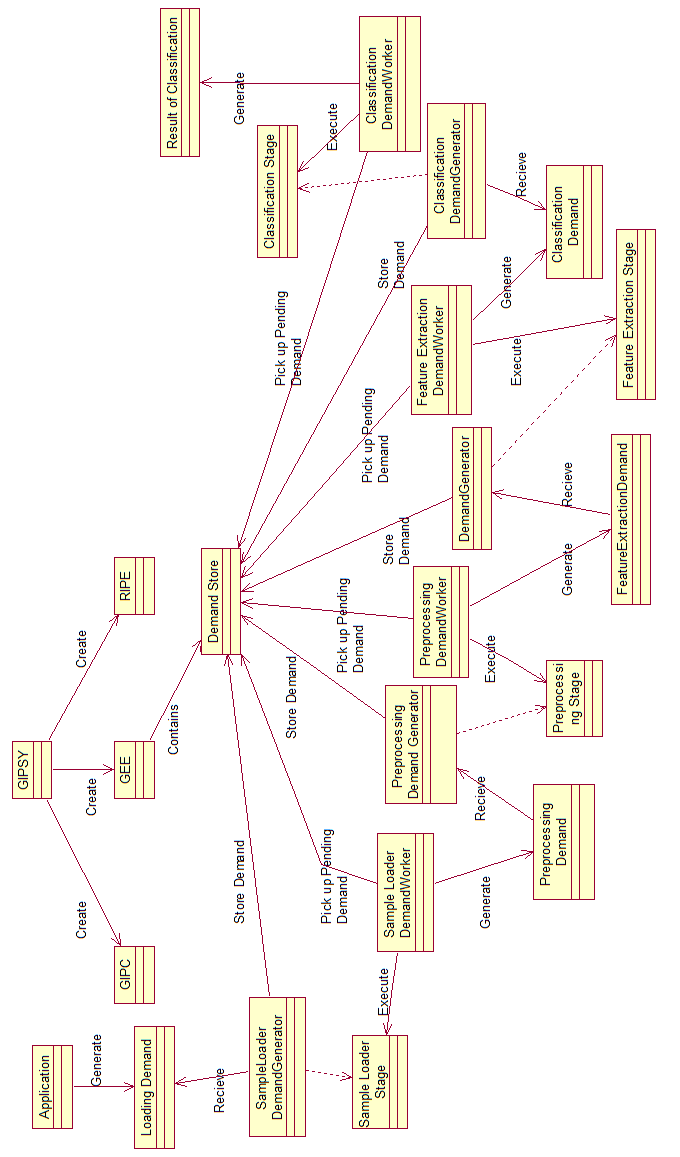}
\caption{Fused DMARF-Over-GIPSY Domain Model}
 \label{fig:D3}
\end{figure*}

\subsubsection{Fused DMARF-Over-GIPSY Run-time Architecture (DoGRTA)} 
 Figure ~\ref{fig:D3} , Shows the fused DMARF-Over-GIPSY domain model. For DMARF to be distributed over GIPSY, discarding communication technology protocols, such as CORBA, RIM and WS, is needed and replacing them with GIPSY$'$s tiers. Basically, GIPSY is a Multi-Tier Architecture which is fully demand driven where the demands are generated by the tiers and migrated to other tiers using the Demand Store by Transport Agent TA [7], [11].

Basically, to get the advantages of the demand driven architecture that GIPSY has, each node of the pipeline has two components, demand generator component and a demand worker component. Generally, the generator is the one who generate the demand. Starting with the sample loader pipeline, the application will create the sample loader pending demand. The demand generator recieves the demand, and store it in the demand store. The sample loader demand worker picks the sample loader pending demand up and carry out the functional computations requested to sample loader stage. Later, the sample loader demand worker generates preprocessing demand. Consequently, the preprocessing demand generator will receives the preprocessing demand and store it in the demand store. The preprocessing demand worker picks up pending or unprocessed preprocessing demand, then it executes the preprocessing stage. After that, the feature extraction demand will be generated by the preprocessing demand worker. The feature extraction demand will be received by the feature extraction generator and stores it in the demand store. The feature extraction worker picks up the feature extraction pending demand and executes the feature extraction stage. Then, the feature extraction worker generates the classification pending demand. The classification generator revieces the classification demand and stores it in the demand store. Then, the classification worker picks up the classification pending demand and executes the classification stage. After that, the classification worker produces a result of the classification stage. The semantic of the MARF$'$s pipelines is maintained by distributing it on a multi-tier architecture like GIPSY. Thus, its scalability is improved [7], [11].

\subsection{Actual Architecture UML Diagrams}

\subsubsection{DMARF}
The entire design of DMARF class diagram is summarized in eight main modules and their relationship as follows: MARF, SpeakerIdentApp, WSUtils, RMIUtils, IsampleLoader, Sample, Ipreprocessing, IFeatureExtraction, and IClassification. Indeed, MARF uses communication technology type of interfaces, WSUtils or RMIUtils, to pick up either manually or automatically which communication technologies is supposed to be in the system design. These interfaces are defined in marf.net and they are used in reflection instantiation utils. Next, RMI, WS server and client interfaces are branched in the hierarchy. They used to set and get in-house-made RemoteObjectReference which isn$'$t a true object reference as in RMI, yet it encapsulates the necessary service location information. These communication technologies are associated somehow with WAL logging and transaction recovery. Also, there are some monitoring modules designed as well. 

Then, SpeakerIdentApp recognizes MARF and commences recognition pipeline. SpeakerIdentApp has several important methods assisting MARF to let sample be identified. Some methods are getVersion, getConfigString, and SetDefaultConfig.  Besides, MARF loads a sample file into IsampleLoader which requires concrete preprocessing to process the sample file after IsampleLoader will ask Sample Class to upload it. Preprocessing will get back to Sample in array list after normalizing has been done. Afterwards, MARF processes this file sample in Preprocessing, and generates preprocessed sample. Likewise, MARF does some further processing with IFeatureExtraction in order to get the sample. 

Furthermore, MARF associates together with IFeatureExtraction to extract features and to generate features vector.  Features which have been generated need to be classified, so MARF and IClassification will address this issue properly. After classifying extracted features has been done, FeatureExtraction get results of array, and determine the feature ID.  Next, it sends this result into Result.  MARF manages this result in both SpeakerIdentApp and Classification then MARF gets this result ID into Result.

        	In Preprocessing Class diagram, it’s summarized into several components (modules) associating with main one, Sample, which are IPreprocessing, IPreprocessingRMI, FFTFilter, IFilter, and Preprocessing.  Sample needs Preprocessing to processes and manages samples using FFTFilter to classify and to preprocess these samples.  In IPreprocessingRMI, it initiates communication technology needed to process this sample as well as to normalize it. 

\begin{figure*} [ht!]
  
  \centering
    \includegraphics[width=0.7\textwidth]{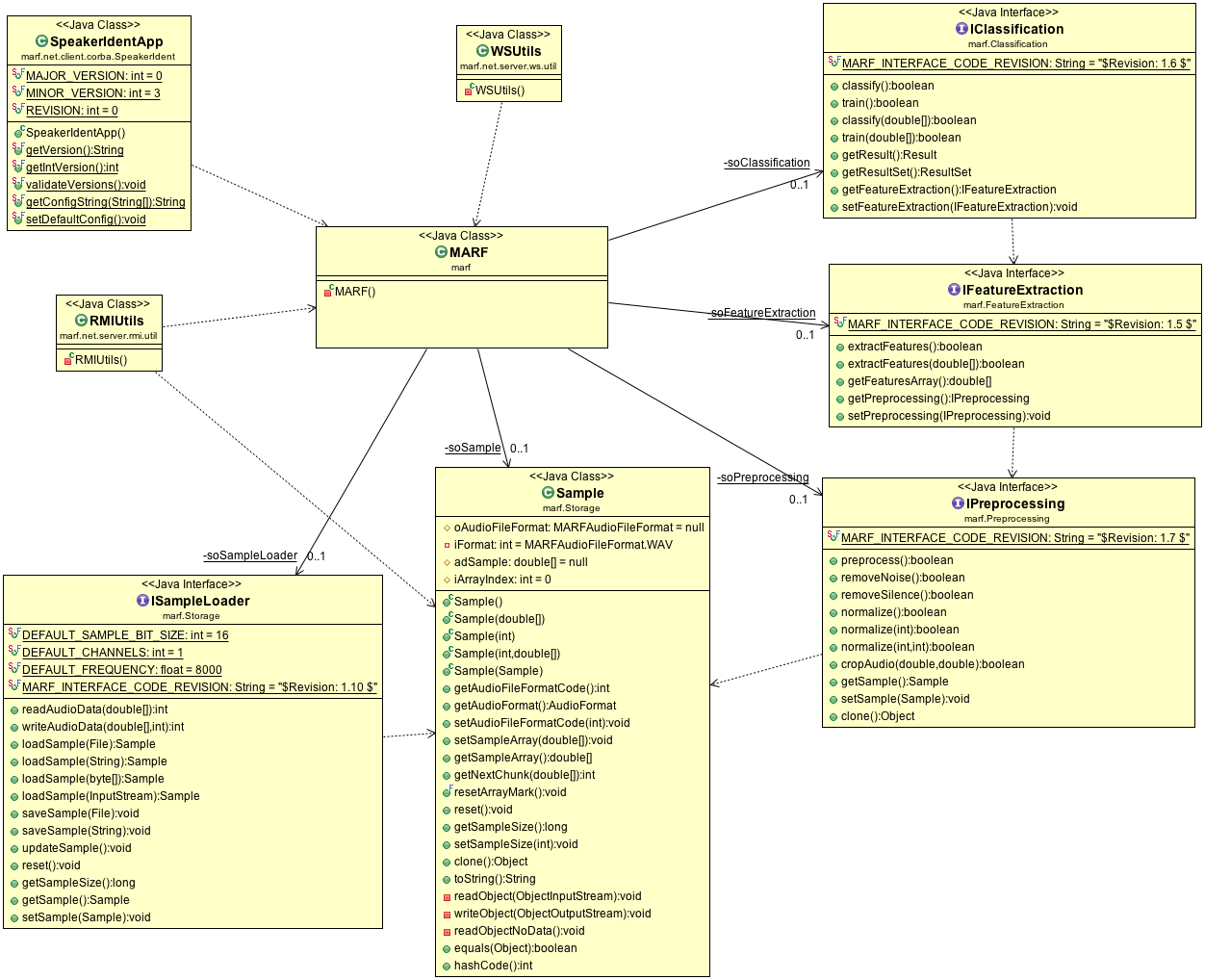}
\caption{DMARF$'$s Class Diagram}
\label{fig:C1}
\end{figure*}

\begin{figure*} [ht!]
  
  \centering
    \includegraphics[width=0.5\textwidth]{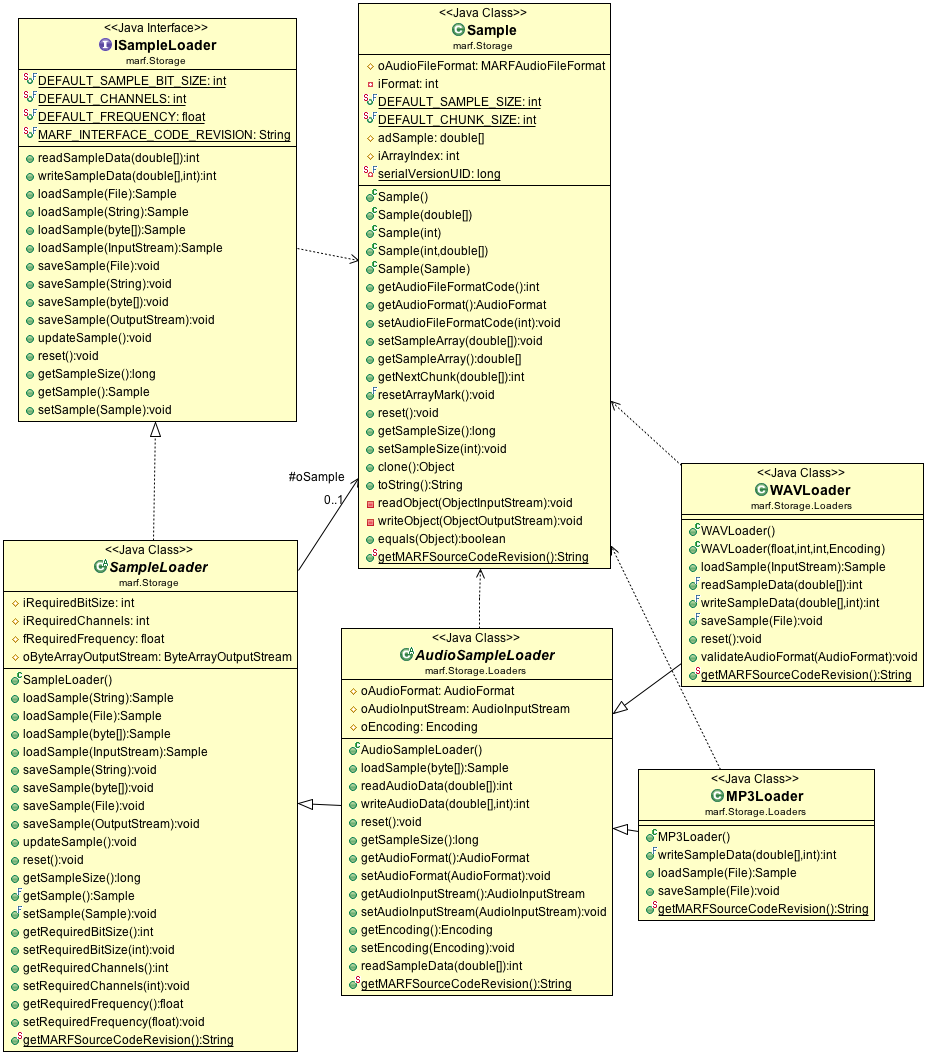}
\caption{DMARF$'$s SampleLoader Class Diagram}
\label{fig:C2}
\end{figure*}

\begin{figure*} [ht!]
  
  \centering
    \includegraphics[width=0.5\textwidth]{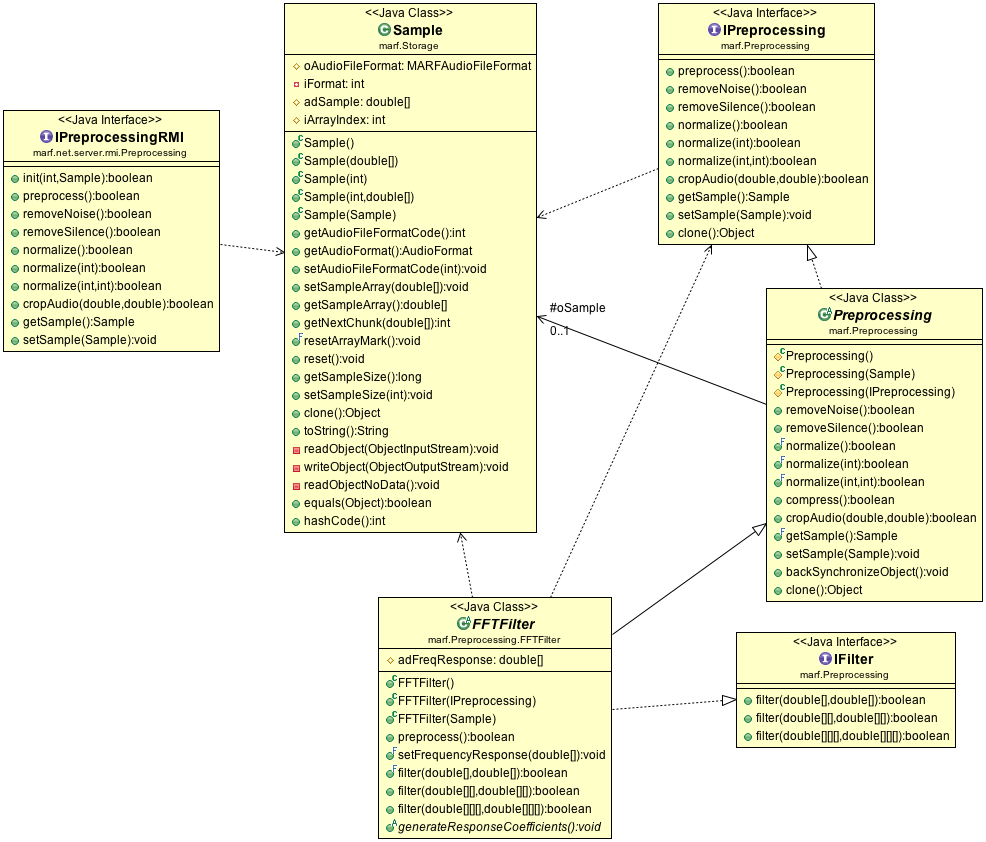}
\caption{DMARF$'$s Preprocessing Class Diagram}
\label{fig:C3}
\end{figure*}

\begin{figure*} [ht!]
  
  \centering
    \includegraphics[width=0.5\textwidth]{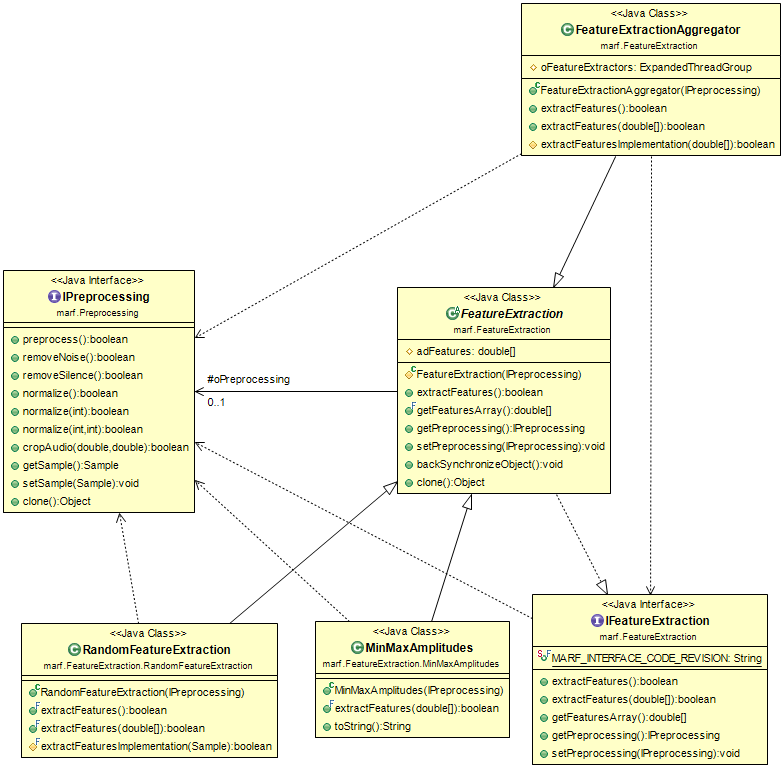}
\caption{DMARF$'$s Feature Extraction Class Diagram}
\label{fig:C4}
\end{figure*}

\begin{figure*} [ht!]
  
  \centering
    \includegraphics[width=0.5\textwidth]{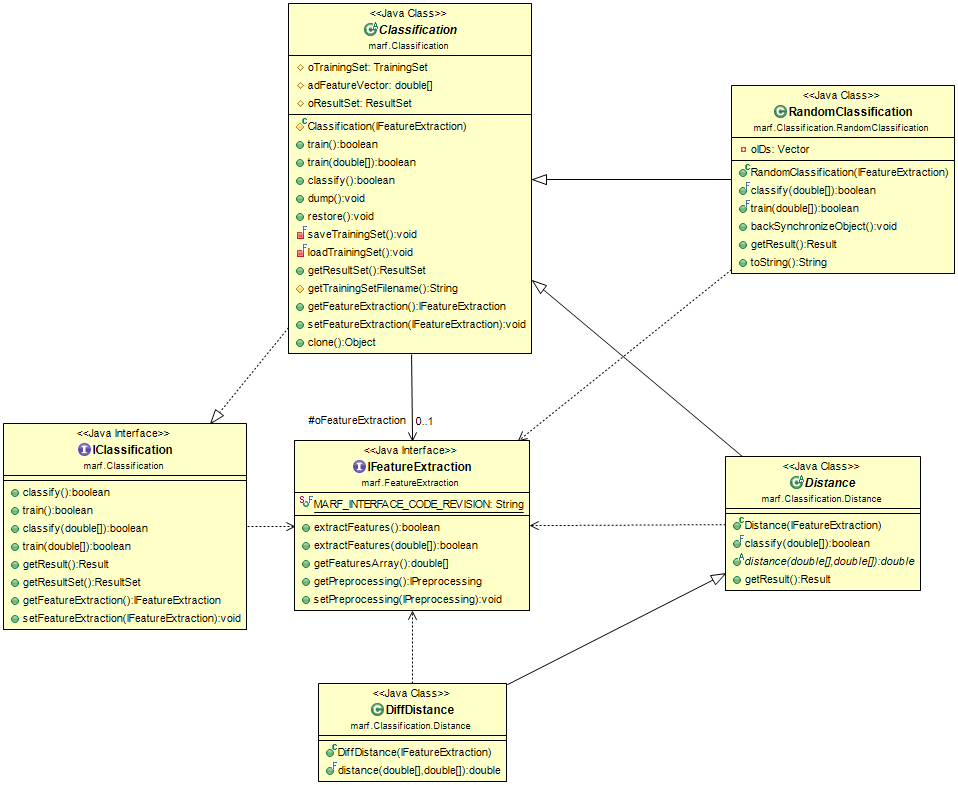}
\caption{DMARF$'$s Classification Class Diagram}
\label{fig:C5}
\end{figure*}

        	Moreover, Sample Loader class diagram gets leveraged from MP3Loader to load sample file as well as to write audio data, to read audio data, and to save a sample.  SampleLoader has loadSample, saveSample, getSampleSize, getSample, setSample, updateSample, and other methods assisting Sample to get audio format, to set audio format, to get next chunk, to reset array mark, to clone, and to get sample array. 
        	In a word, DMARF class diagram stems from DMARF’s domain model with some solution problem models which express how MARF address and process its samples in an order and management to have good results. Not only does MARF identifies a user sample, but also it classifies, preprocesses it and extracts features. Communication technologies such as RMI, WS have been used for communication with WAL, managed his serialized WAL entity handle by storage (Storage Manager) and transaction, interacts with delegate type. The StorageManager class provides implementation of serialization of classes in binary as well as compressed binary formats. Not only that, but It also has facilities to plug-in other storage or output formats. In terms of Database, all result set, and classifications states which is written in SpeakerIdentApp are stored in database [2]. Figures ~\ref{fig:C1}, ~\ref{fig:C2}, ~\ref{fig:C3}, ~\ref{fig:C4} and ~\ref{fig:C5}, show the DMARF$'$ class diagrms.

DMARF$'$s domain model (Conceptual architecture) covers most world problems. Components such as MARF communicate with samples through the communication technology to distribute services. 
Once a  sample has been uploaded in Survey System by the  end-user, SpeakerIdentApp will identify the user depending on DMARF. A user  could record his voice. In contrast, the actual system architectures shows the internal components and the interactions between them. For instance, in the former a voice has represented in Sound Sample whereas in the latter is represented as Sample.

From our conceptual classes, most of the conceptual classes maps to the actual classes in DMARF class Diagram, and they are as follow:
 \begin{itemize}
\item  
Speaker Ident App maps to SpeakerIdentApp
 
\item 
Voice Sample maps to Sample
\item DMARF maps to DMARF

\item MARF maps to MARF$'$s SampleLoader, Preprocessing, FeatureExtraction, Classification
\end{itemize}

Conceptual classes uses classes from the actual system architecture. Since DMARF deals with pattern recognition, the conceptual classes uses the pattern recognition and its components to authenticate the user in it’s domain problem. Conceptual classes such as SpeakerIdentApp, SoundSample, etc uses corresponding similar classes from the actual classes. This means, the conceptual system is fully dependent on the actual DMARF architecture and its components. In the conceptual architecture, the voice sample uploads to DMARF through the SpeakerIdentApp and DMARF is communicated through the communication technology.  Therefore, the conceptual diagram is more of an application level architecture of DMARF. While the DMARF actual architecture explains about the components and interaction between the components.  

We have used ObjectAid UML tool to help us build the class diagram for both DMARF and GIPSY. ObjectAid UML is a plugin for Eclipse that is used to reconstruct the class diagrams for the actual system source code. This plugin allows to reverse engineer the class diagrams just by drag and drop of the required java classes from the source code. Also, it allows to show the associations and dependencies between different classes. In addition, we have the option to hide/show different types of attributes and methods.

  \begin{figure} [ht!]
  
  \centering
    \includegraphics[width=0.5\textwidth]{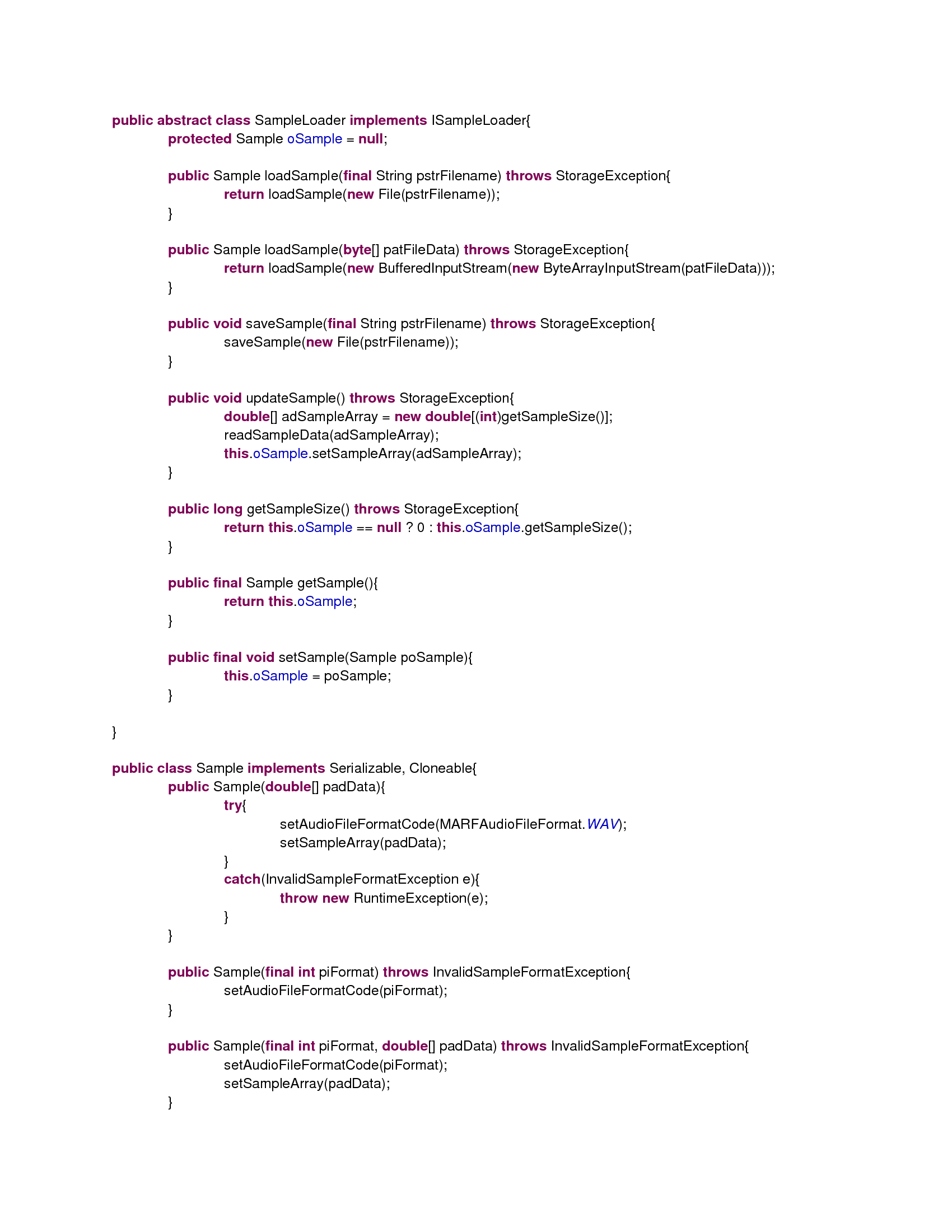}
\caption{DMARF$'$s class diagram Classes and the Relationship Source Code First Part.}
 
\end{figure}

 \begin{figure} [ht!]
  
  \centering
    \includegraphics[width=0.5\textwidth]{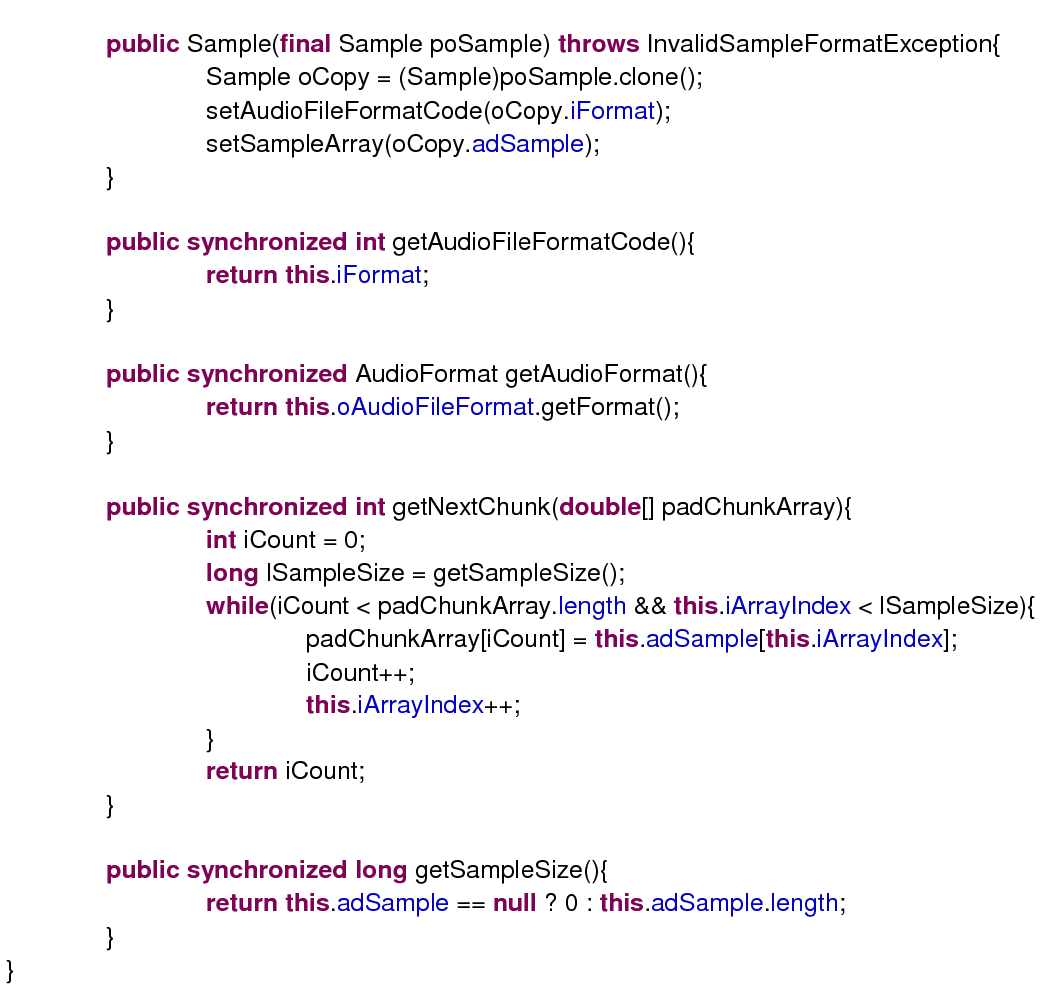}
\caption{DMARF$'$s class diagram Classes and the Relationship Source Code Second Part.}
 
\end{figure}

\subsubsection{GIPSY}

 \begin{figure*} [ht!]
  
  \centering
    \includegraphics[width=0.5\textwidth]{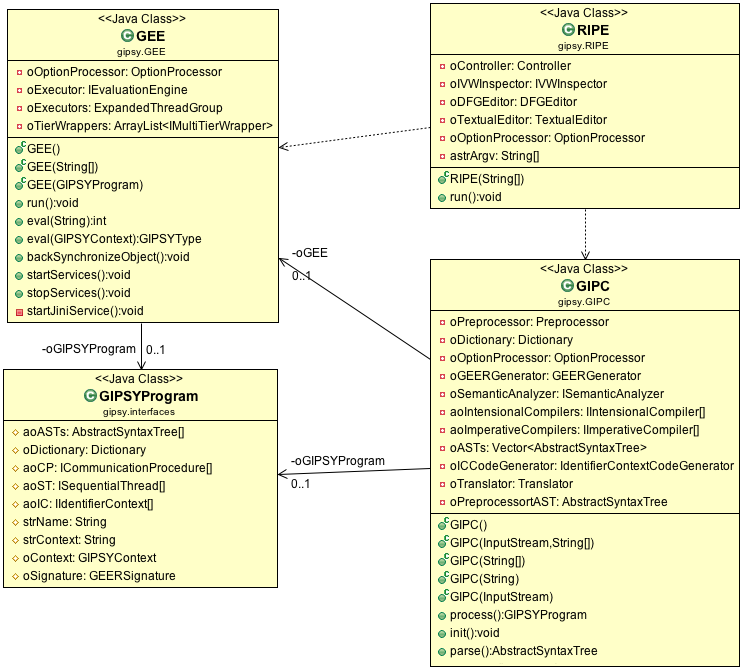}
\caption{GIPSY$'$s Class Diagram}
\label{fig:C6}
\end{figure*}

 \begin{figure*} [ht!]
  
  \centering
    \includegraphics[width=0.7\textwidth]{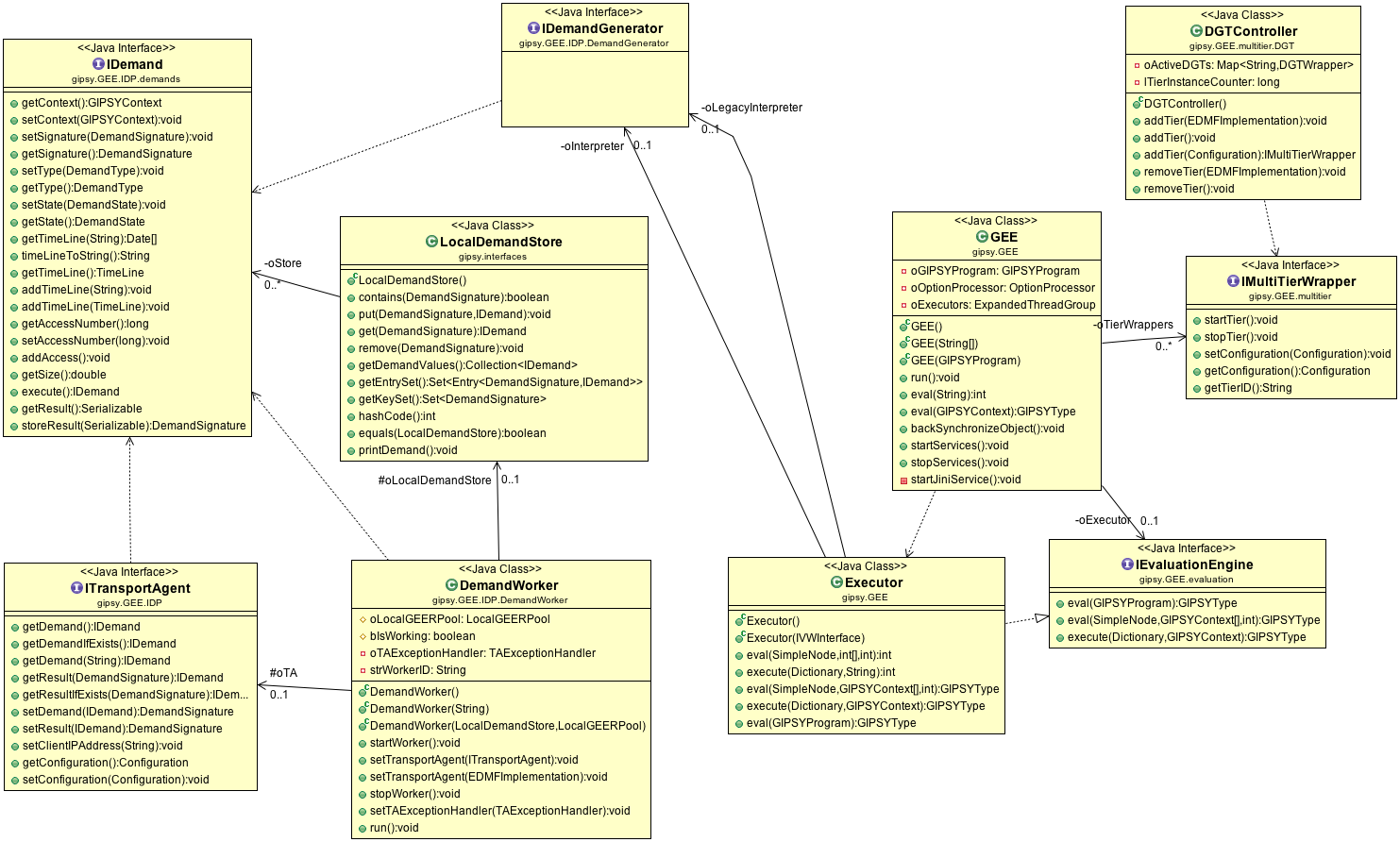}
\caption{GIPSY$'$s GEE Class Diagram}
\label{fig:C7}
\end{figure*}

\begin{figure*} [ht!]
  
  \centering
    \includegraphics[width=0.5\textwidth]{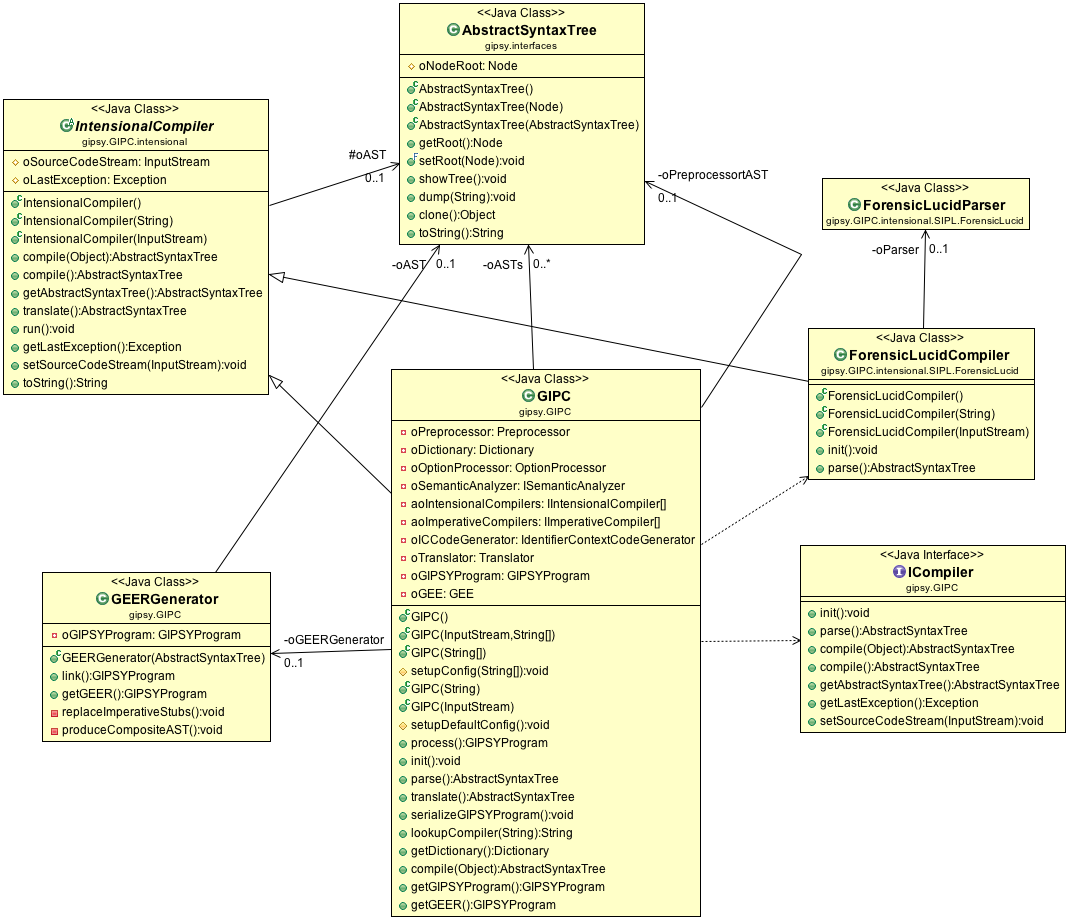}
\caption{GIPSY$'$s GIPC First Class Diagram}
\label{fig:C8}
\end{figure*}

\begin{figure*} [ht!]
  
  \centering
    \includegraphics[width=0.5\textwidth]{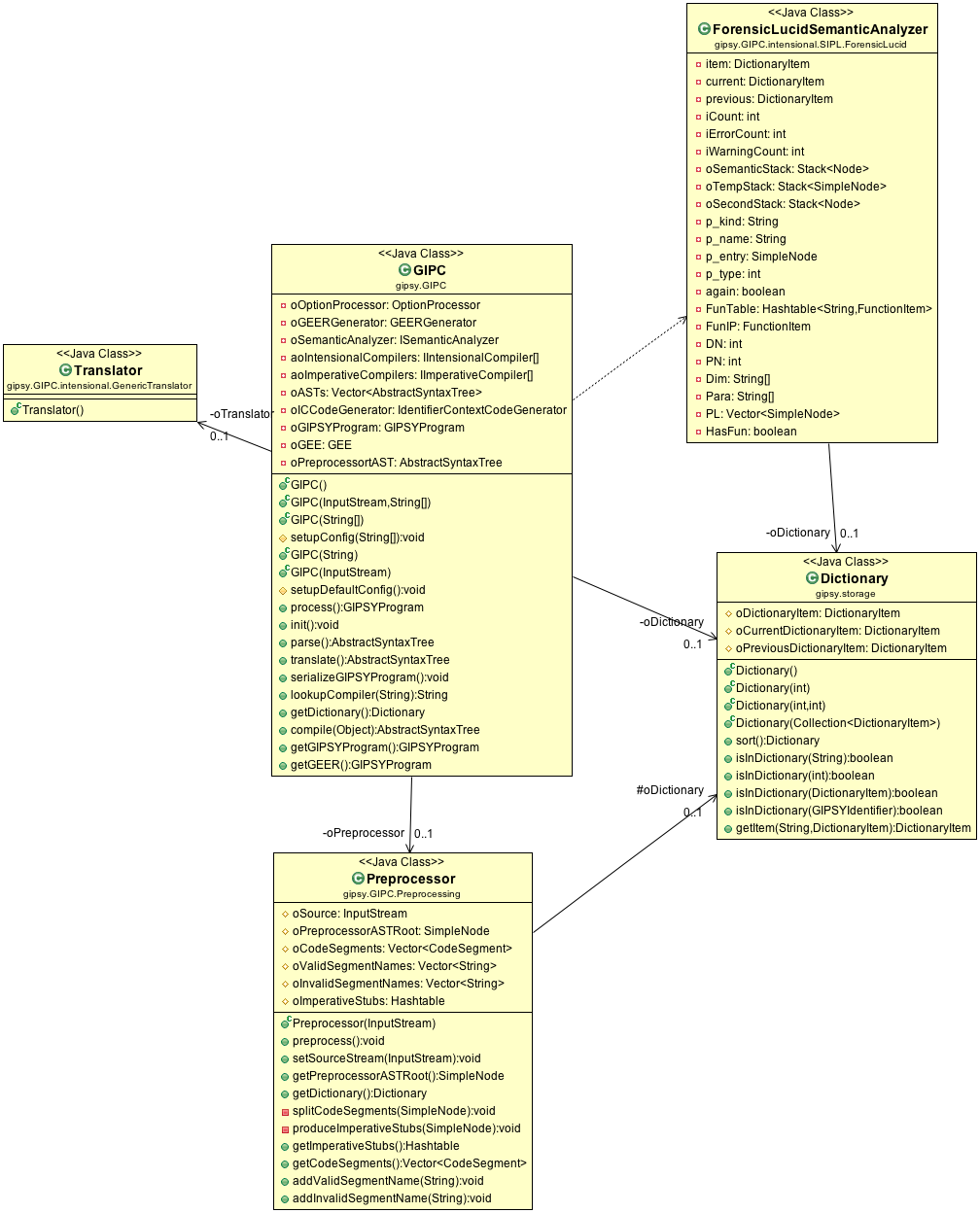}
\caption{GIPSY$'$s GIPC Second Class Diagram}
\label{fig:C9}
\end{figure*}

The whole design of GIPSY class diagram contains major modules and some their relationships between these modules: GIPSYProgram, GIPC, GEE, RIPE and other important components. Several design patterns have been applied such as observer, and singleton pattern. The singleton is applied to ensure one controller is created for each tier on each node, and further factory method pattern lets the subclasses to specify the tier objects it controls. We abbreviate unimportant classes; we focus only the interested classes which relative into our domain model. GIPSYProgram has dictionary, communication procedure, and GEERSignature associate with other main components GEE and GIPC. GEE class is integrated to invoke multi-tier services via the main entry of the engine and to accommodate missing development.  GEE$'$s controller will adjust and process services over a particular tier. To let GEE start, GEE uses created API. 
        
	Next, GEE employs factory method to instantiate desired tier type of arbitrary instances, and then it starts or stops them directed using API. GEE delegates some activities to node controller and proceeds to the program execution if there is one to execute after having a tier get started.  As execution processes, demands are generated and handed into the tier that responsible for the delivery and results computation to get back with warehouse store so that it caches principles.

GIPC has Preprocessor, Dictionary, Translator, and others that cooperate together to investigate Intentional Programming Language, requiring GEE, RIPE Conroller, DFG editor, and Abstract Syntax Tree in GIPSYProgram. GIPC uses GEERGenerator to link GIPCProgeam so Abstract Syntax Tree could clone, dump, and sow tree. GIPC has compiler to parse, compile, and get abstract syntax tree. ForensicLucidCompiler is one type of GIPC compiler has ForencisLucidParser.

 	In addition, GEE has major components: DemandGenerator, LocalDemandStore, DemandWorker, and TransportAgent. LocalDemandStore is considered as an observer for its subjects, DemandGenerator which gets or sets state, and gets or sets access for DemandWorker object who interests to know about DemandGenerator new status which informing the concerete subject, TransportAgent, to return subject state [7].  

The conceptual architecture is an application of the the actual system architecture of GIPSY. That is, the conceptual model uses the GIPSY architecture to function certain tasks. In the conceptual architecture, GIPSY is used to compile the FingerPrint Application and it is communicated through the communication technology. Therefore, the conceptual diagram is more of an application level architecture of GIPSY. Whereas, the GIPSY actual architecture explains about the components and interaction between the components. Figures ~\ref{fig:C6}, ~\ref{fig:C7}, ~\ref{fig:C8}, and ~\ref{fig:C9}, show the GIPSY$'$ class diagrms.

From our conceptual classes there are only two classes that maps to actual classes in GIPSY class Diagram, and they are as follow:

 \begin{itemize}
\item  
GIPSY maps to GIPSYProgram, GIPC, GEE, and RIPE.
\item  
Communication technology maps to RMI, and CORBA.
 
\end{itemize}

The actual classes obtained from the background reading and the actual architecture of the system are closely associated with the conceptual classes. Some conceptual classes which comes in the application level apart from the GIPSY implementation has some discrepancy between them, whereas the conceptual classes for the GIPSY has similar actual classes. That means, the actual architecture of the system has similarity with the conceptual architecture when it comes to the interaction between the GIPSY$'$s components.

\section{Methodology}

\section{Refactoring}

\subsection{Identification of Code Smells and System Level Refactorings}

\subsubsection{DMARF}

After inspecting the DMARF codes, we come across with the following code smells:

In the {\bf MARF class}, we found three code smells namely code duplication, Long method and complex if statements, and they are as follow:
  \begin{itemize}
\item  
train () and train (Sample poSample) - Code duplication:
In both the methods, some codes are repeated, which causes code duplication. In this case, we could extract the duplicated codes from both the methods and make it as a method, so that it can be called from both these methods.
\item  
startRecognitionPipeline(Sample poSample) - Long method:
This particular code smell is due to large chunk of codes in the specified method which makes it hard to understand. To avoid the code smell, we could extract the codes from this long method and create another methods and call it as needed. 
 
\item  
startRecognitionPipeline(Sample poSample) - Complex if statement:
This method contains complex if statements and it may lead to longer processing time and logical errors. So the possible solution is to simplify the if conditions. 
\end{itemize}

In the classes {\bf Distance and RandomClassification}, one code duplication code smell is found. 
getResult() - Code duplication:

  \begin{itemize}
\item  
Both the Distance and RandomClassification classes have the same superclass Classification class. Both classes have the getResult() method with similar behavior. This means that the method is duplicated content and to solve this problem is to pull up this method to the superclass and inherit its behavior. 
\end{itemize}

In the classes NeuralNetwork, one code duplication code smell is found.
generate() - Code duplication:
\begin{itemize}
\item  
In this method, some codes are repeated, which makes code duplication. Therefore, we could extract the codes from this method and make different methods and call it from the original method. 
\end{itemize}

Another code smell found among the following classes is speculative generality. It is affected by the classes AIFFCLoader, AIFFLoader, AULoader, MIDILoader, MP3Loader, SNDLoader and ULAWLoader. These classes are not implemented but just put it as empty. The possible way to avoid is to remove these Loader classes. 

\subsubsection{GIPSY}

Based on the analysis made on the GIPSY package,  we found the following code smells:

In {\bf GEE class}, three methods are affected by the code smell long method:
\begin{itemize}
\item  
GEE(String[] argv) - Long method:
This method has long codes which are prone to logical errors or complex structure. So the possible solution is to extract the codes and make corresponding methods and call them whenever wanted. 
\item
startServices() - Long method:
Similar to the above method, it also has long codes which are prone to logical errors or complex structure. So the possible solution is to extract the codes and make corresponding methods and call them whenever wanted. 
\item
eval(GIPSYContext poContext) - Long method:
This method also has long codes which are prone to logical errors or complex structure. So the possible solution is to extract the codes and make corresponding methods and call them whenever wanted. 
\end{itemize}

Similarly In the GIPC class, we found one possible code smells, which are:
GIPSYProgram process() - Long method:

\begin{itemize}
\item 
This method is very long, open to errors and has complex code structure that makes it hard to read and understand. So the possible solution is to extract fragment of codes and make them separated methods and call it as needed. 
\end{itemize}

In the JavaCompiler class, we found two code smells:
\begin{itemize}
\item 
init() and parse() - Code duplication:
These two methods are inherited from the ImperativeCompiler class and has similar codes. So the possible solution is to extract the codes from both these classes and make a single method in the super class and call it as needed. 
\end{itemize}

As we found in the DMARF, we also found some Speculative Generality code smells in the GIPSY package as well. The following classes (PerlCompiler, CCompiler, CPPCompiler, FortranCompiler, PythonCompiler) are not fully implemented. So we could remove them from the code to avoid any possible confusion or errors.

\subsection{Specific Refactorings that Will Be Implemented in PM4}

\subsubsection{DMARF}
We identified the common design problems or code smells in DMARF$'$s source code and their corresponding refactoring techniques among the methods, and they are as follows: 
 
\begin{itemize}
\item 
checkSettings() Method from MARF class - Complex if statement 

\item 
startRecognitionPipeline(Sample poSample) Method from MARF class - Long method

 \item generate() Method from NeuralNetwork class - Code duplication

\item 
startRecognitionPipeline(Sample poSample) Method from MARF class - Complex if statement

 \end{itemize}

For DMARF, there is no test cases for NeuralNetwork, Distance Class, RandomClassification Class and Classification class. Therefore, in this case we can write the JUnit tests manually. Eclipse supports the creation of JUnit tests via a wizard. Naming the new JUnit tests would be TestNeuralNetwork, TestDistance, TestRandomClassification and TestClassification, like the TestWaveLoader test case. We will write these tests to make sure that the code behaviour did not change after the refactoring. For MARF class, there is test case called test under default package.

\subsubsection{GIPSY}
For the refactoring in GIPSY, we will try to refactor the code smells in the classes GEE, GIPC and JavaCompiler. For the class GEE, we will try to refactor the methods GEE(String[] argv) and startServices() has long method code smell. In addition, in GIPC class, we encounter a long method code smell for the method GIPSYProgram process() and we will try to refactor it too. Regarding to the class JavaCompiler, we will refactor the code duplication code smells for the methods init() and parse() by pulling them up to ImperativeCompiler.

For the testing in GIPSY, we have the gipsy.tests.Regression class to apply the test cases to see if the refactoring makes any bad changes to the output of the system. This test class is mainly useful for the GEE and GIPC class code smell refactorings. There is no test cases for JavaCompilerclass in GIPSY$'$s source code, but if we may to create JUnit test, it would be TestJavaCompilerclass, like the TestForensicLucidSemantic test case. These test classes will help us to identify if the refactoring has affect the behaviour of that class or not. If the outcome of the test cases is similar before and after each refactoring, then we could say that the refactoring was effective and did not affect the outcome of the system. If the outcome is different from the outcome before refactoring, then we could conclude that the refactoring badly affected the system and should make a revoke of the doings and give more attention to the refactoring. 

\subsection{Identification of Design Patterns}

In order to understand and identify design patterns, we have used ObjectAid UML Explorer tool. ObjectAid UML Explorer is optimized for the quick and easy creation of UML class from existing Java source code and libraries. It uses the UML notation to show a graphical representation of existing code. ObjectAid UML plugin for Eclipse is used to find out or reverse engineer the interacting classes from the actual classes. Based on the  Design-Pattern Identification, the corresponding interacting classes are identified using this eclipse plugin [27].

\subsubsection{DMARF}

\begin{figure*} [ht!]
  
  \centering
    \includegraphics[width=0.5\textwidth]{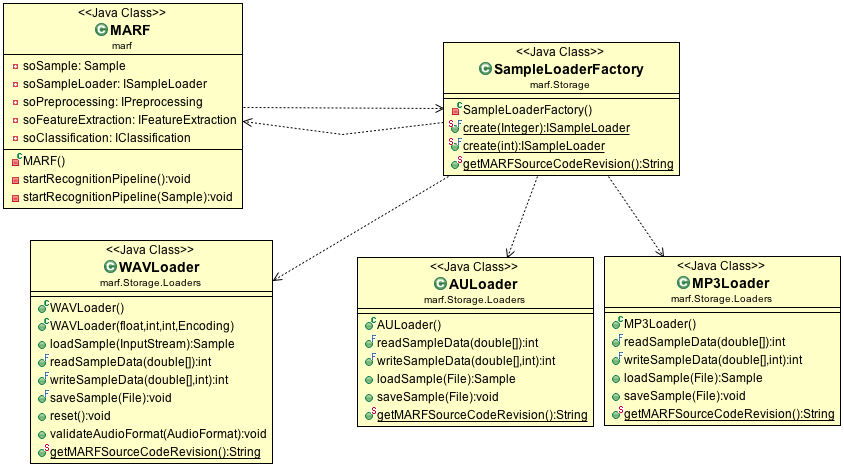}
\caption{DMARF$'$s classes involved in the Factory pattern.}
 \label{fig:DP1}
\end{figure*}

\begin{figure} [ht!]
  
  \centering
    \includegraphics[width=0.5\textwidth]{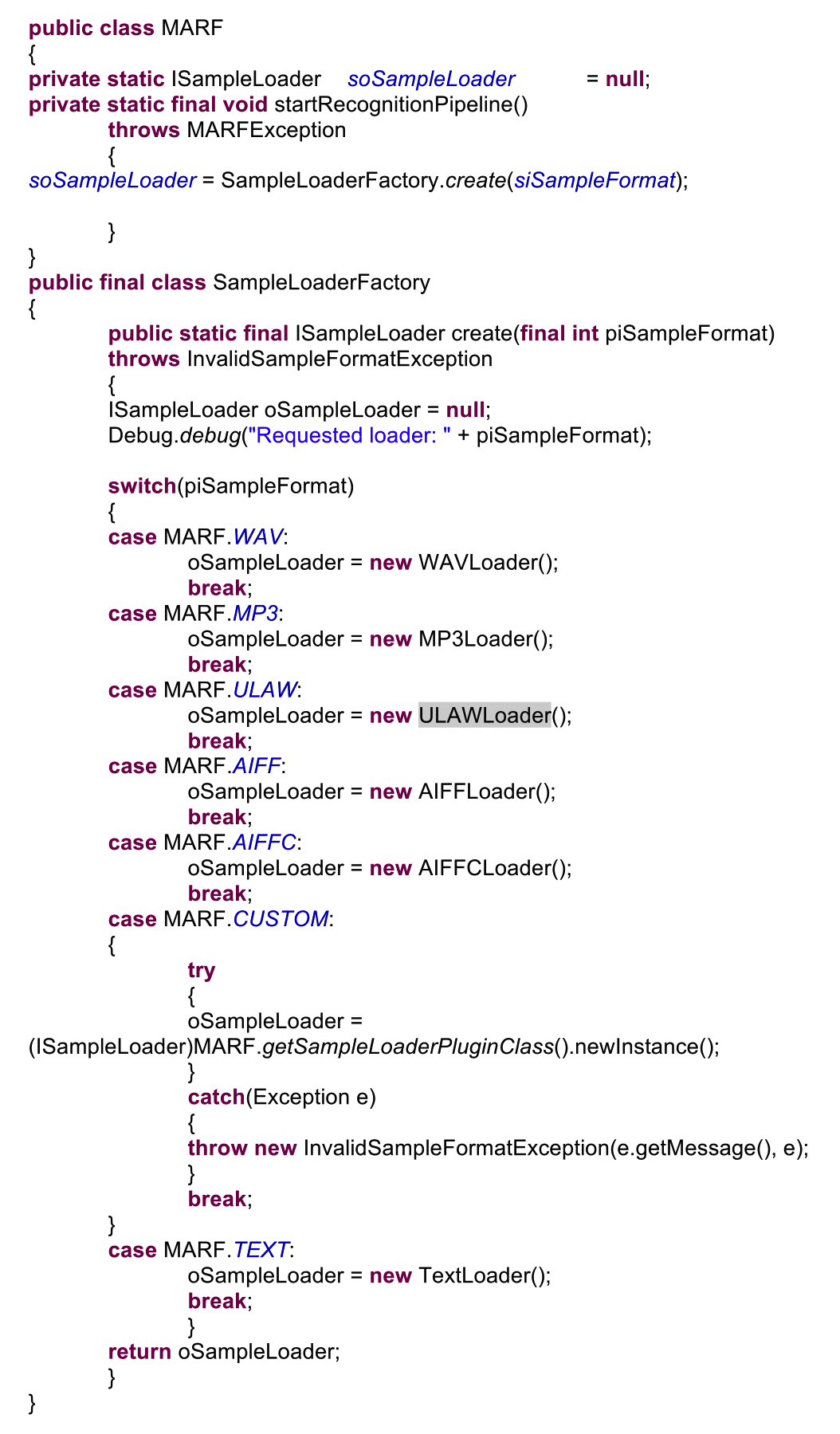}
\caption{DMARF$'$s Factory Pattern Code}
 \label{fig:DP2}
\end{figure}

\begin{figure*} [ht!]
  
  \centering
    \includegraphics[width=0.5\textwidth]{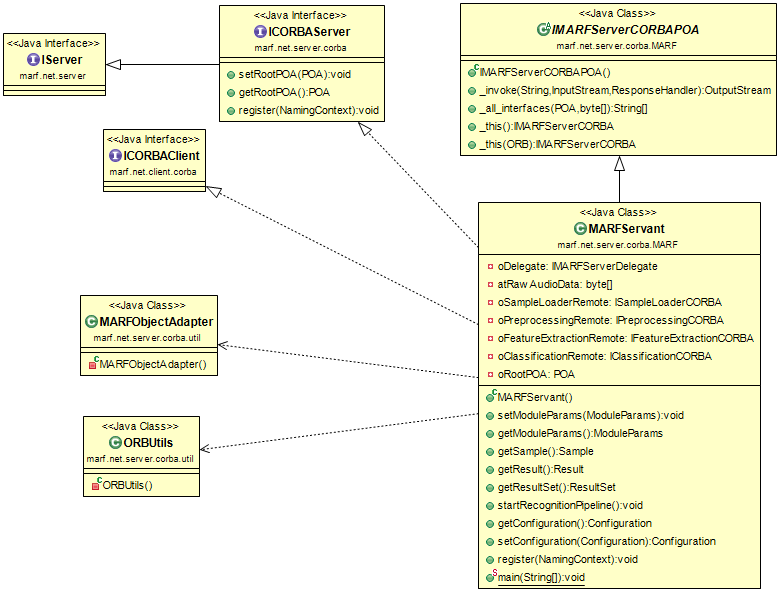}
\caption{DMARF$'$s classes involved in the Adapter pattern.}
 \label{fig:DP3}
\end{figure*}

\begin{figure} [ht!]
  
  \centering
    \includegraphics[width=0.5\textwidth]{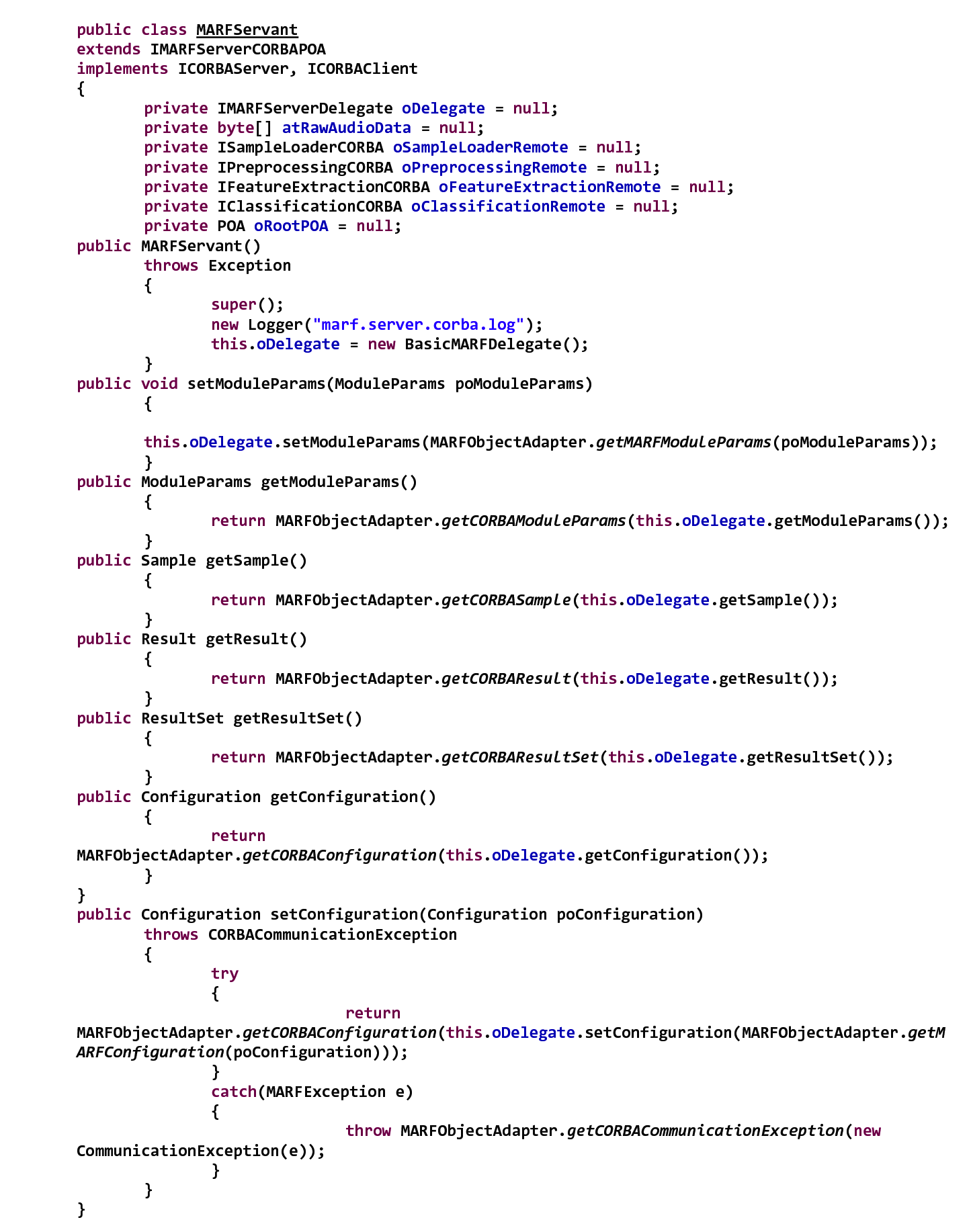}
\caption{DMARF$'$s Adapter Pattern Code First Part}
 \label{fig:DP4}
\end{figure}

\begin{figure} [ht!]
  
  \centering
    \includegraphics[width=0.5\textwidth]{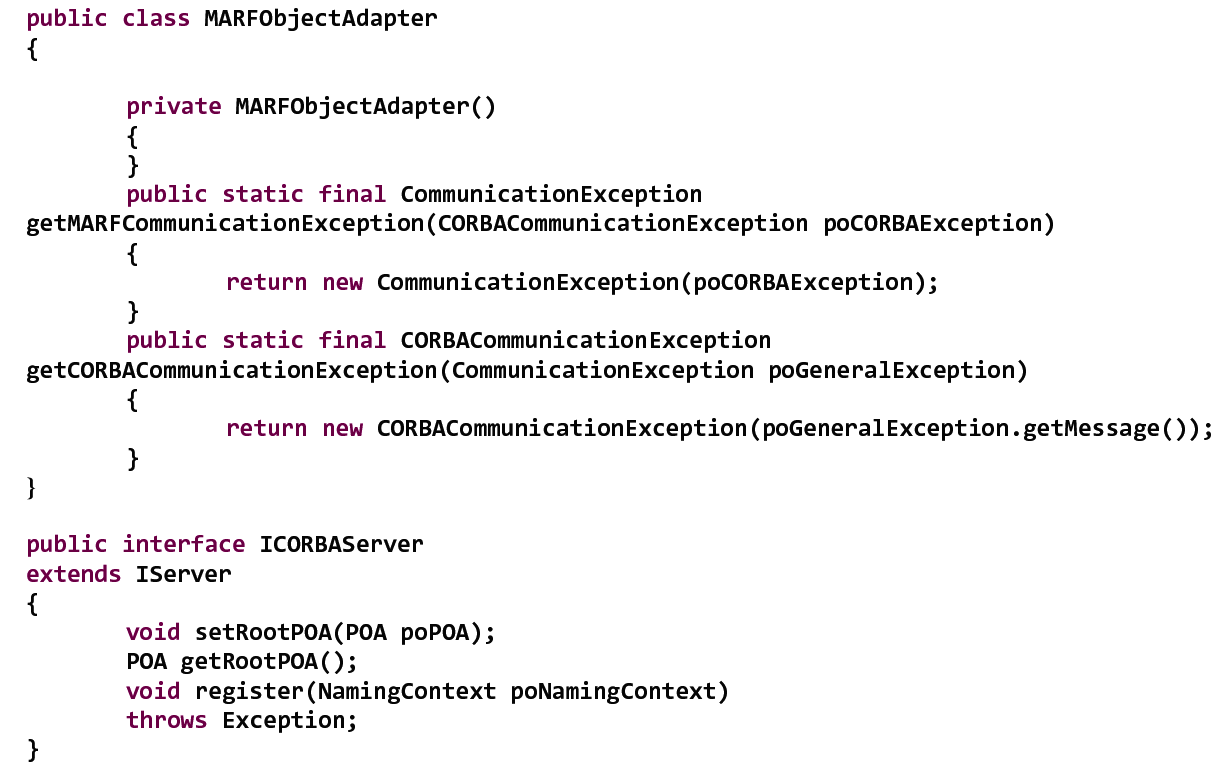}
\caption{DMARF$'$s Adapter Pattern Code Second Part}
 \label{fig:DP5}
\end{figure}

\begin{figure} [ht!]
  
  \centering
    \includegraphics[width=0.4\textwidth]{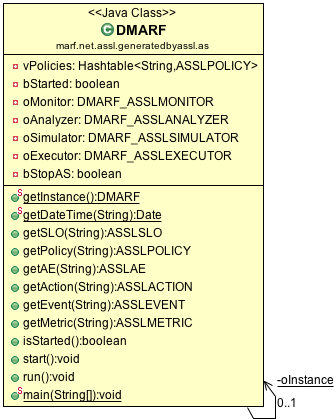}
\caption{DMARF$'$s classes involved in the Singleton pattern.}
 \label{fig:DP6}
\end{figure}

\begin{figure} [ht!]
  
  \centering
    \includegraphics[width=0.5\textwidth]{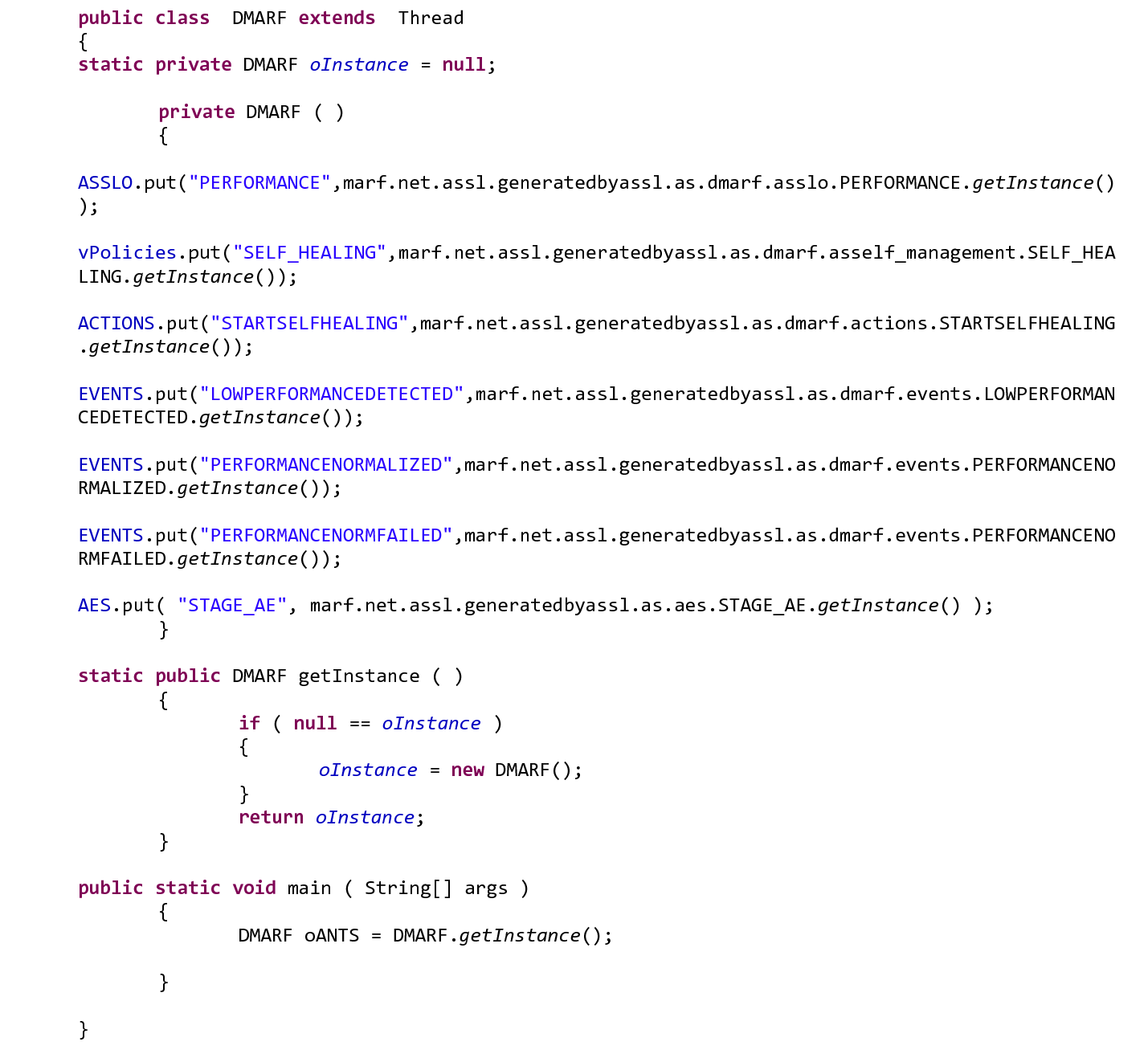}
\caption{DMARF$'$s Singleton Pattern Code}
 \label{fig:DP7}
\end{figure}

      \mbox{} \\
\noindent
{\bf Factory Pattern:}

The factory pattern is a widespread design pattern. This pattern comes under creational pattern, and it helps to create an object in the best way by providing varies way to create it. This pattern comes in handy when there is a need to create an object without directly specifying the type of the object at compile time, and allows the client to select the desired class at runtime to create the object. To implement the factory pattern, an interface class has to be created and be the only place where these objects can be instantiated. Client classes select the type of object they need, and send it to the factory class. Then, depending on the sent type of the needed object, the factory class creates the object, and returns it back to the client class [22].
 
Factory pattern is used in DMARF$'$s sampleloader pipeline stage to support loading different voice extensions, like WAVE, MP3 and many more. A SampleLoaderFactory class is implemented to allow MARF class to select the type of the sample to be loaded by SampleLoader class. Then, it instantiate the selected type of object and returns it back to MARF class to load that sample. Figure ~\ref{fig:DP1}, shows DMARF$'$s classes involved in the factory pattern. Figure ~\ref{fig:DP2}, shows source code for factory pattern in DMARF.

Sometimes, creating objects is a complex process, and if it were not solved probably, it could cause lots problems in the code. Especially, when the needed object is unknown before run time. Code duplication would have been used to solve this problem. However, it is not the right way to do it. The best way for this kind of problem is to use the factory pattern [22].

  \mbox{} \\
\noindent
{\bf Adapter Pattern:}

The adapter design patterns replicates the plug adapter, in that it converts the current systems interface into another interface by wrapping the entire interface and creating the desired new interface. In the other words Adapter design pattern is one of the structural design pattern and its used so that two unrelated interfaces can work together. The object that joins these unrelated interface is called an Adapter. The benefits of this pattern are that it allows objects to be encapsulated by a new class structure and creates new interfaces that match the class that invokes it [23]. There are two approaches  whereas  implementing Adapter pattern :  class adapter which mean uses java inheritance and extends the source interface, and object adapter which mean uses Java Composition and adapter contains the source object , however both these approaches produce same result [24].

In DMARF, the adapter pattern is applied while implementing the CORBA services, a data type adapter had to be made to adapt certain data structures that came from MARF.idl to the common storage data structures. Thus, the MARFObjectAdapter class was provided to adapt these data structured back and forth with the generic delegate when needed. In the other words, MARFObjectAdapter class is responsible for translating common MARF data structures to CORBA and vice versa. Figure ~\ref{fig:DP3}, shows a UML class diagram for the adapter pattern in DMARF. Figure ~\ref{fig:DP4} and ~\ref{fig:DP5}, shows source code for adapter pattern in DMARF.

 \mbox{} \\
\noindent
{\bf Singleton Pattern:}

Singleton pattern is one of the simplest design patterns that addressed these concerns: create only one instance of a class, allow a global point of access to the object, and  allow multiple instances in the future without affecting a singleton class$'$s clients[ 22].

Singleton pattern is needed in DMARF to provide a global visibility or a single access point to a single instance of the class DMARF, in this case, always a unique instance of DMARF class will be instantiated to avoid the synchronization problems that can be raised. The constructor is assigned to be private, in order to prevent direct instantiations from other clients, and to define other subsystems based on this extending . The static method  getInstance ( ), is to ensure that a single instance of DMARF object is returned,  it allows the instantiation of DMARF elements just when the singleton instance is equal to null. Figure ~\ref{fig:DP6} , shows a UML class diagram for the singleton Pattern in DMARF. Figure ~\ref{fig:DP7}, shows source code for singleton pattern in DMARF.

\subsubsection{GIPSY}

\begin{figure*} [ht!]
  
  \centering
    \includegraphics[width=0.5\textwidth]{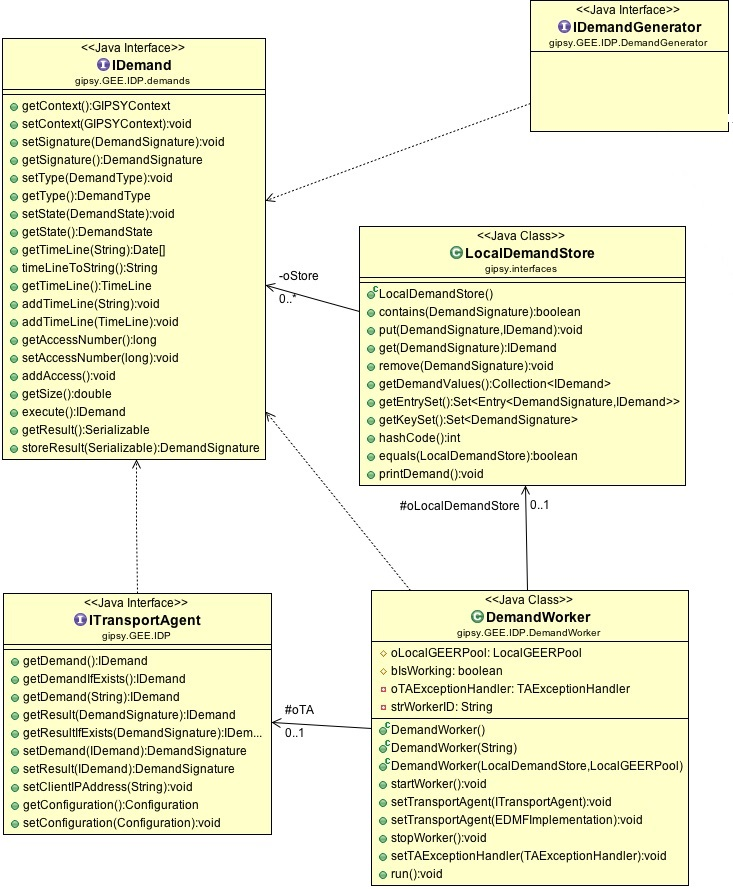}
\caption{GIPSY$'$s classes involved in the Observer pattern.}
 \label{fig:DP8}
\end{figure*}

\begin{figure} [ht!]
  \centering
    \includegraphics[width=0.5\textwidth]{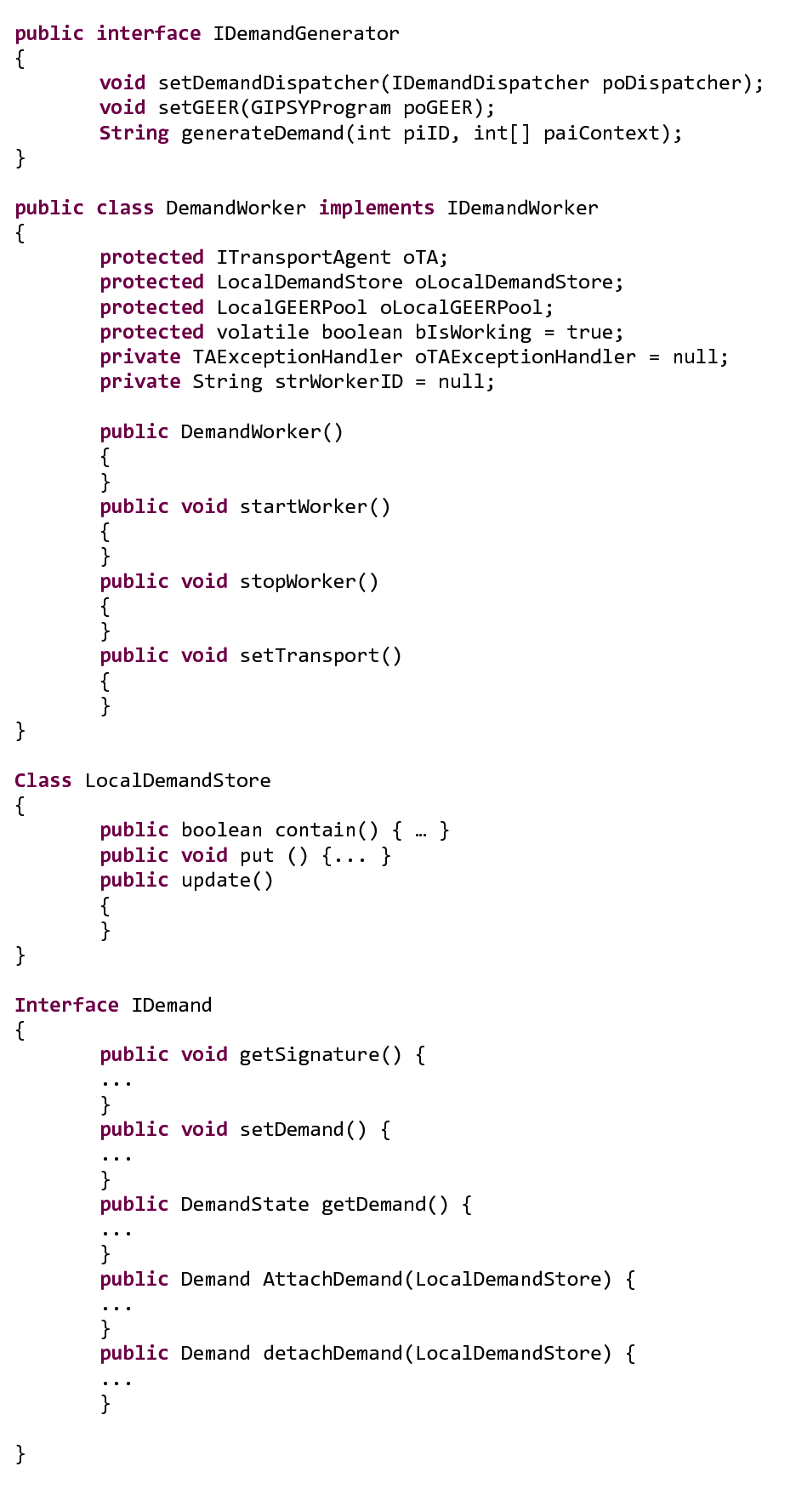}
\caption{GIPSY$'$s Observer Pattern Code First Part}
 \label{fig:DP9}
\end{figure}

\begin{figure} [ht!]
  \centering
    \includegraphics[width=0.5\textwidth]{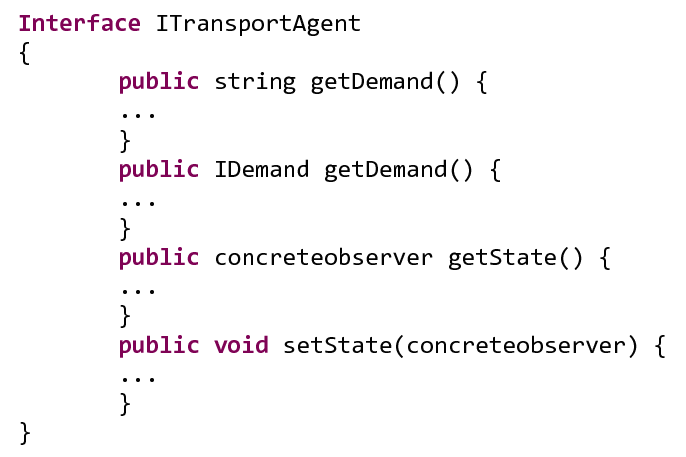}
\caption{GIPSY$'$s Observer Pattern Code Second Part}
 \label{fig:DP10}
\end{figure}

 \begin{figure*} [ht!]
  
  \centering
    \includegraphics[width=0.5\textwidth]{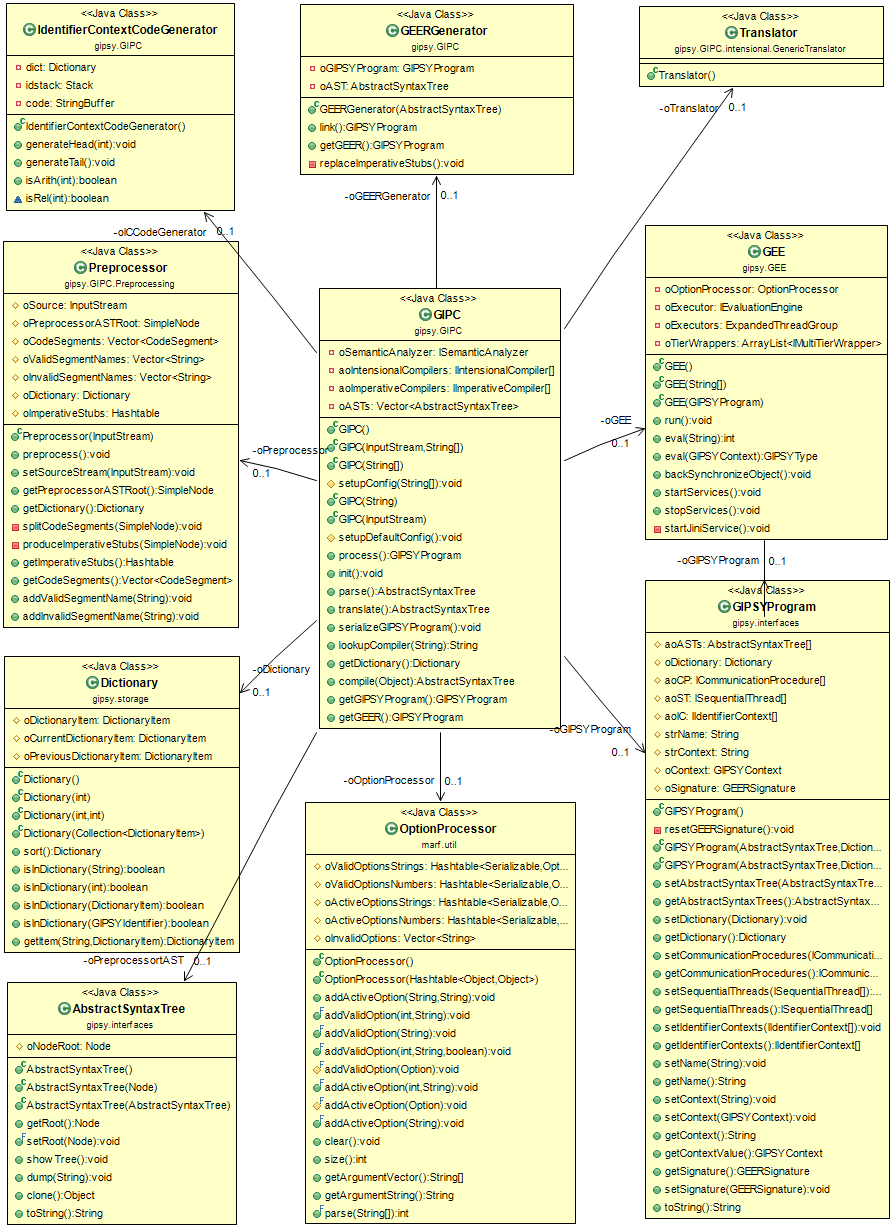}
\caption{GIPSY$'$s classes involved in the Facade pattern.}
 \label{fig:DP11}
\end{figure*}

\begin{figure} [ht!]
  
  \centering
    \includegraphics[width=0.5\textwidth]{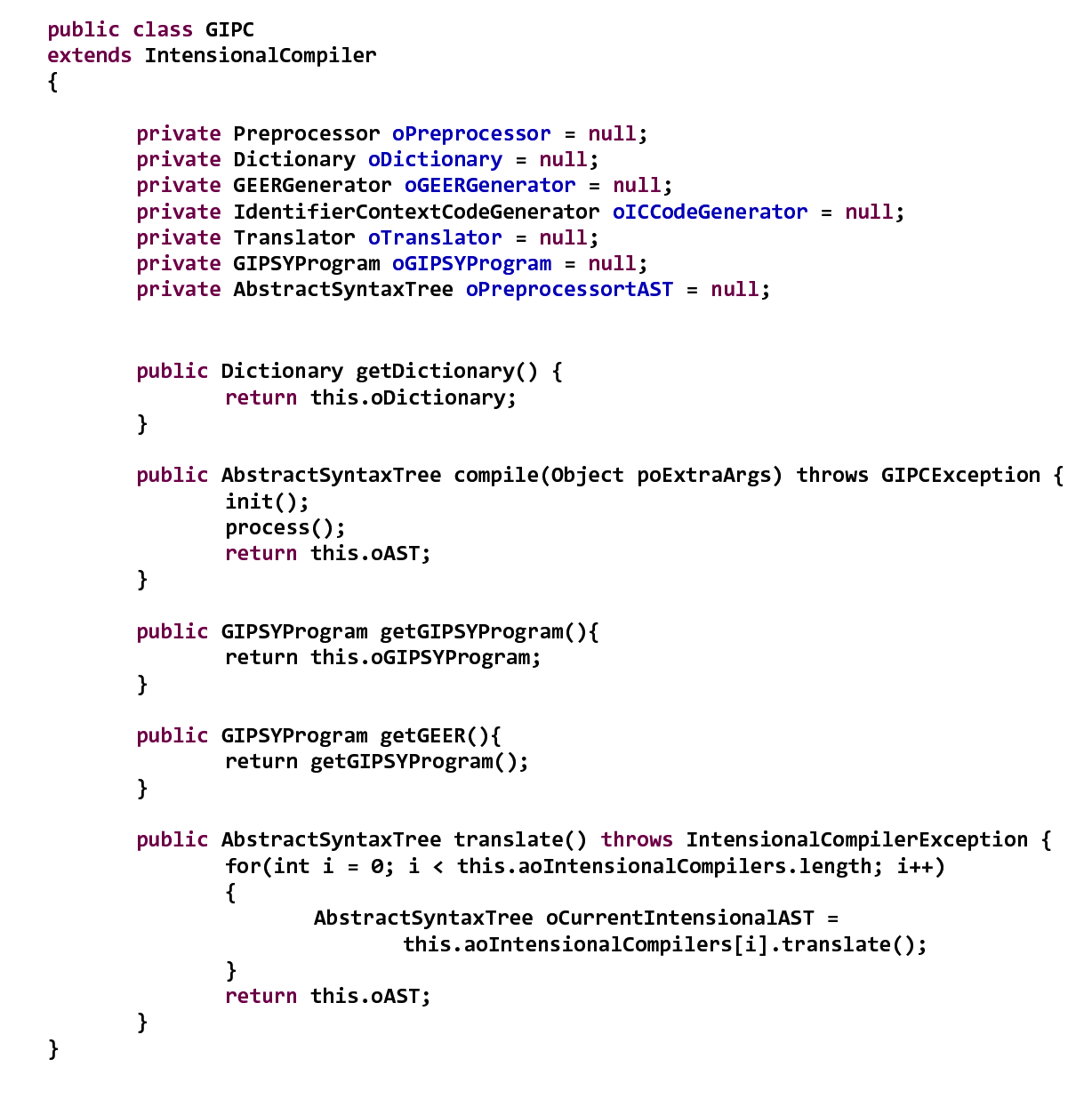}
\caption{GIPSY$'$s Facade Pattern Code}
 \label{fig:DP12}
\end{figure}

\begin{figure*} [ht!]
  
  \centering
    \includegraphics[width=0.5\textwidth]{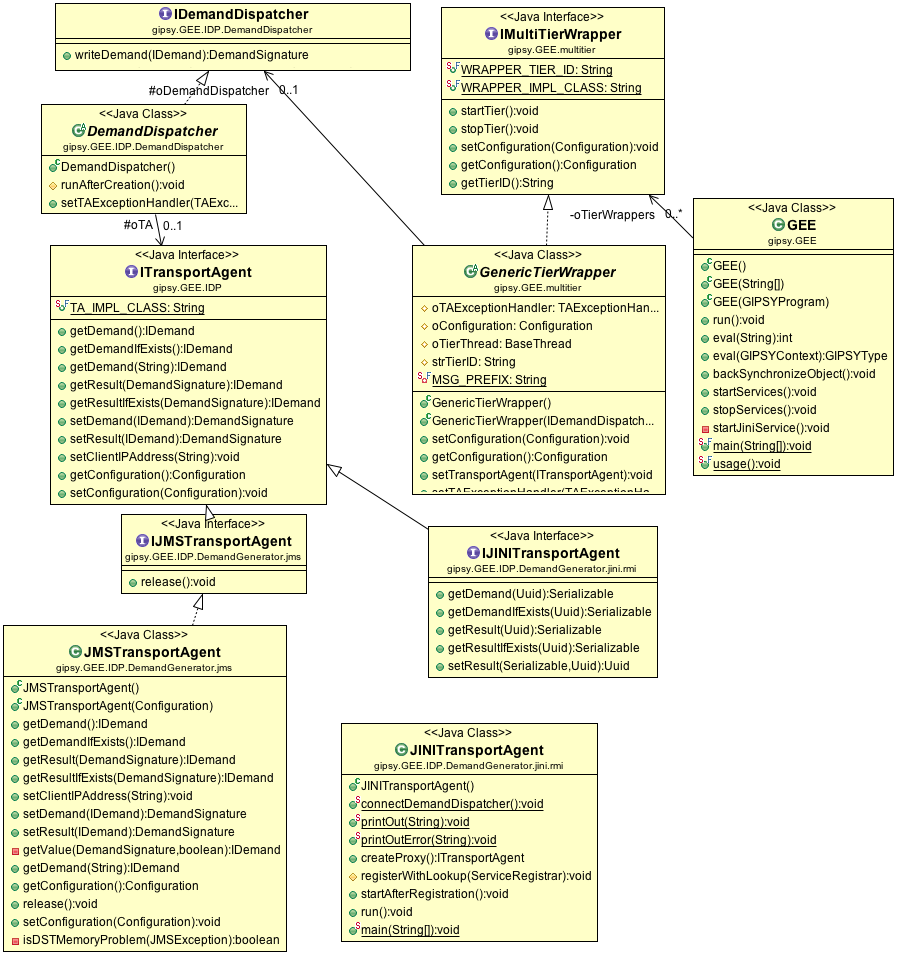}
\caption{GIPSY$'$s classes involved in the Strategy pattern.}
 \label{fig:DP13}
\end{figure*}

\begin{figure} [ht!]
  \centering
    \includegraphics[width=0.5\textwidth]{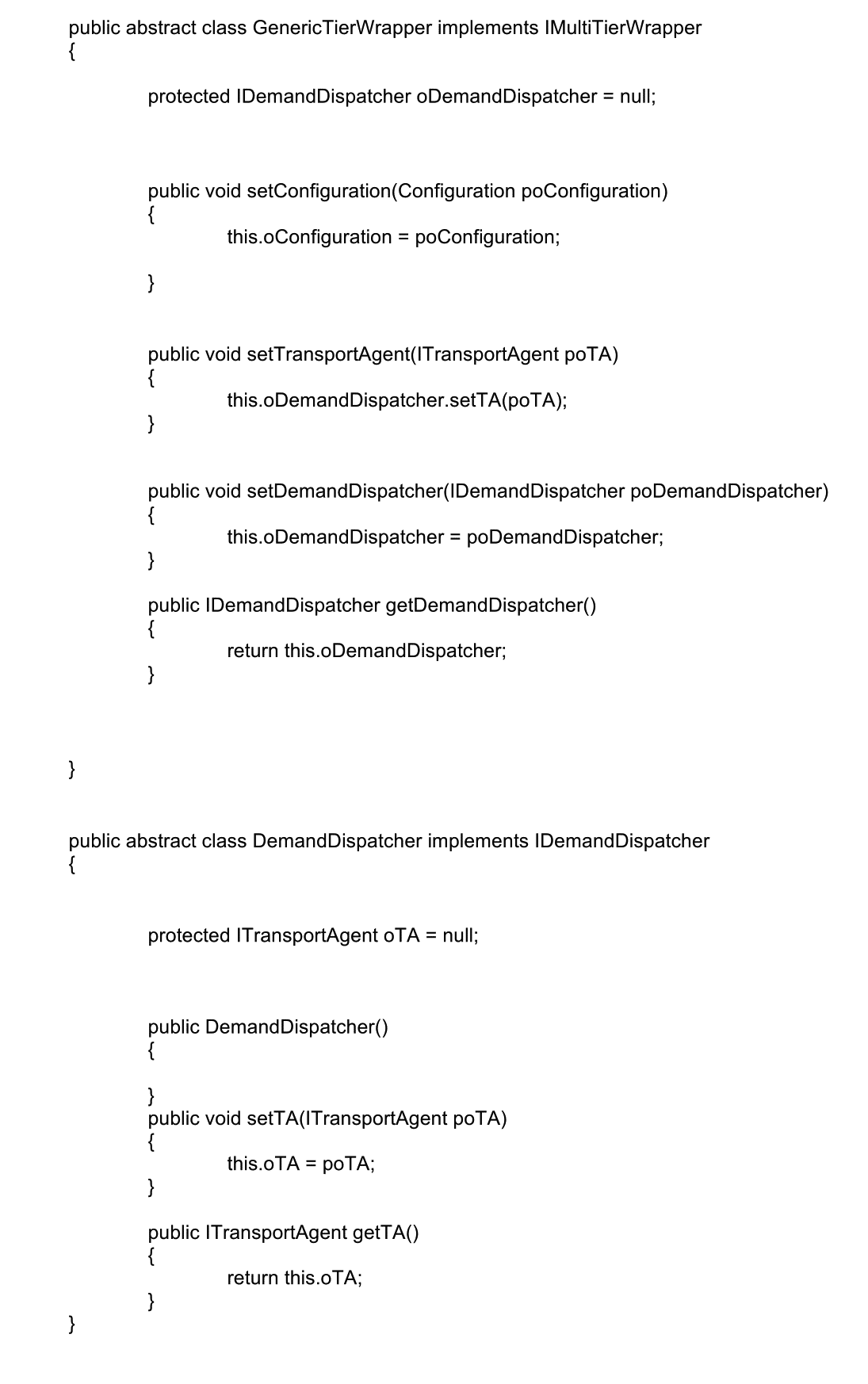}
\caption{GIPSY$'$s Strategy Pattern Code}
  \label{fig:DP14}
\end{figure}

 \mbox{} \\
\noindent
{\bf Observer Pattern:}

Observer pattern is known as publish subscribe. Define a one to many dependency between object so that when one object changes state, all its dependents are notified and updated automatically. The need to maintain consistency between related objects is a common side-effect of partitioning a system into a collection of cooperating classes [22].  

Using the observer pattern could be suitable once: an abstraction has two aspects one dependent on the other; a change to one object requires changing others and you do not know how many objects need to be changed; or an object should be able to notify other objects without making assumptions about who these objects are.  
 
Observer pattern has couple participants assisting and cooperating to each other in order to achieve their target. Subject knows its observers, and provides an attaching and detaching interface observer object. Observer defines an updating interface for object that should be notified. Concrete subject stores state of interest to concrete subject object. Concrete observer maintains a reference to concrete subject object. 

GEE$'$s observer pattern separates the presentation of user interface from underlying application data is one of aspect which could be reused independently. Once Demand generator (a subject) gets executed and changed its state, LocalDemandStore (observer), contains demand signature, will notify Demand Worker (concrete observer) his new state; thereby TransportAgent (concrete subject) will returns subject state. This behavior implies that Demand generator and RransportAgen are dependent on the data object, and get notified of any changes in its state. Indeed, observer pattern describes how to establish these relationship. All observers are informed whenever the subject undergoes a change in state and synchronize its state as well. Figure ~\ref{fig:DP8} , shows a UML class diagram for the observer pattern in DMARF. Figure ~\ref{fig:DP9} and  ~\ref{fig:DP10}, shows source code for observer pattern in GIPSY.

 \mbox{} \\
\noindent
{\bf Facade Pattern:}

The facade pattern provides an interface for a large amount of code and makes it easier to use the code library. It reduces the dependencies outside the code library and hides the implementation of the subsystems from any clients and makes it easy to use. It also makes the code more easier to read and understand. So this structuring reduces the complexity of the subsystem. These subsystems can be any groups of classes. [22]

 Figure , facade is interacting with the different classes and GIPC acts as an interface for all these classes. GIPC calls the methods from the classes like GIPSYProgram, Preprocessor, GEERGeneration, Translator within the main class. So the developer could easily interact with the methods of the other classes without interfering with the underlying code. It can also call the other class methods and attributes using the facade. Figure ~\ref{fig:DP11} , shows a UML class diagram for the facade Pattern in GIPSY. Figure ~\ref{fig:DP12} , shows source code for facade pattern in GIPSY.

 \mbox{} \\
\noindent
{\bf Strategy Pattern:}

The strategy pattern define sets of algorithms where each set is encapsulated in order to be interchangeable. This pattern allows algorithms to differ independently from clients that use it. More importantly, this pattern retains the open/closed and reuse principle [25].

In GIPSY this pattern is applied where there are two implementation options can be set to the TransportAgent, Jini and JMS. All TAs are implementing the ITransportAgent interface;  the implementations of the two DMSs, Jini and JMS are encapsulated in IJMSTransportAgent and IJiniTransportAgent which in turn implement the ITransportAgent hence they are interchangeable. The main class GEE or any external applications can invoke IMultiTierWrapper interface, which is implemented by the abstract class called GenericTierWrapper. In the GenericTierWrapper class, the TransportAgent will be set by through the demand dispatcher. This technique allows the implementation of  transport agents vary independently from clients that use it [26].Figure ~\ref{fig:DP12} , shows a UML class diagram for the strategy Pattern in GIPSY. Figure ~\ref{fig:DP13}, shows source code for strategy pattern in GIPSY.

\section{Implementation}
\subsection{Refactoring Changesets and Diffs}

\subsubsection{DMARF}

In DMARF, we had done three code refactorings, in order to reduce its complexity to be more understandable and easy to maintain. After making these changes, we made two test cases to make sure that the DMARF’s behavior did not change. We created  a test case called TestMARF to
test checksettings() method to ensure that its behavior isn$'$t  effected by the refactoring, and for startRecognitionPipeline(), we used default.test class to test the behavior of the method. Also, we have created TestNeuralNetwork to test generate() method behavior. Fortunately, all the tests showed that the refactoring did not affect the behavior of the methods.     

\begin{itemize}
\item 
Change 1/4: Exract methods from startRecognitionPipeline(Sample poSample) from MARF class in DMARF:

This has more than one task, where it should run the whole MARF’s pipeline stages. As a result, this method is really long, and it will not be easy to comprehend the method from the first look to the code. Therefore, we extracted this method to three methods, where each one runs different stage.

\item 
Change 2/4:  Reduce if statement complexcity in checksettings() from MARF class in DMARF:

In checksettings() method there is a complex if statement, which could be not easy to understand on first sight. In order to reduce its complexity, we extracted all the condition from the if statement, and created new variables with a naming that can tell the code reader about its meaning. These variables carry the result of the old conditions.      

\item 
Change 3/4:    Extract method from generate(int piNumOfInputs, int[] paiHiddenLayers, int piNumOfOutputs) from Neural Network class in DMARF:

This method has code duplication when creating different layers for the system. In order to get rid of  the code duplication, we had to combine the three layers code into one check point and also we broke down this long method into helper and sub methods. We simplified if statement and for loop.
\end{itemize}

\subsubsection{GIPSY}

In GIPSY, we only made one code refactoring. This was made on Process() method, which was really long and impossible to understand it at one sight. After extracting couple of methods from Process(), we tried the gipsy.tests.Regression test to make sure that the behavior did not change. Fortunately, the tests showed that the refactoring did not affect the behavior of the method.    

\begin{itemize}
\item 
Change 4/4: Exract methods from Process() method from GIPC class in GIPSY:

This method suffers from long method code smell and complex structure, and it is doing more than one task. Therefore, it would be really hard to understand or maintain this method. In order to reduce its complexity, we broke it down into couple of methods and separate each different task in a different method.
\end{itemize}
\section{Conclusion}

One of the major problems in software development is the poor design and coding. This directly affects the quality of the software system under development. One of the best methodology to follow while coding is to follow the design patterns. Every pattern deals with a specific problem in the design or implementation of a software system. They help us to include the existing well proven coding methods in software development which follows a good design methodology. GIPSY$'$s and DMARF$'$s source codes are designed using many patterns such as Factory, Observer, Singleton, Adapter, Facade and many others. This helps to construct the software architecture with specific properties. Here, the purpose of implementing GIPSY is to have a flexible compiler architecture that could read and compile multiple intentional programing languages, whereas DMARF helps to match and recognize different patterns. In order to get more clarification, we created the domain model of two systems which uses GIPSY and DMARF namely Survey System and Crime Investigation System. 

During the code review, we come across some poor decay of design, called code smells. This can be eliminated using the code refactoring techniques. It cleans up the code in a systematic and efficient manner. In the GIPSY and DMARF projects, we observed some code smells such as code duplication, long method, feature envy and speculative generality. These code smells are then evaluated and solved using different strategies such as method extraction and pull up method. In order to make sure that the behaviour of the system isn’t changed, we conducted JUnit test-cases during the refactoring. The final code after the refactoring is much reliable and efficient compared to the non-refactored code with code smells.

\bibliographystyle{IEEEtran}
\bibliography{bibliography}


%\clearpage

%\section*{Paper distribution}

 \begin{table}[htpb]
    
  \renewcommand{\arraystretch}{1.3}
 \caption{ Paper distribution: First Deliverable} 

    \begin{tabular}{|c|c|p{2cm}|}

\hline
Name                &       & Paper Title                                                                                                                             \\ \hline
 \vtop{\hbox{ \strut Abdulrhman } \hbox{ \strut Albeladi }}& DMARF & \vtop{\hbox{ \strut  {[}2{]}On design and implementation of} \hbox{ \strut distributed modular audio recognition} \hbox{ \strut framework: Requirements and specification } \hbox{ \strut design document.}} \\ \cline{2-3} 
                    & GIPSY & \vtop{\hbox{ \strut { [}11{]} Using the General Intentional} \hbox{ \strut Programming System (GIPSY) for} \hbox{ \strut evaluation of higher-order intentional logic} \hbox{ \strut  (HOIL) expressions.}}  \\ \hline

\vtop{\hbox{ \strut Aber } \hbox{ \strut Abozkhar }}      & DMARF &\vtop{\hbox{ \strut  {[}3{]}Managing distributed MARF} \hbox{ \strut with SNMP  }}                                                                                            \\ \cline{2-3} 
                    & GIPSY & {[}9{]}The GIPSY architecture                                                                                                           \\ \hline
\vtop{\hbox{ \strut Ahmed} \hbox{ \strut Almessabi}}     & DMARF &\vtop{\hbox{ \strut {[}5{]}Towards a self-forensics} \hbox{ \strut property in the ASSL toolset }}                                                                           \\ \cline{2-3} 
                    & GIPSY & \vtop{\hbox{ \strut {[}8{]}GIPSY a platform for the investigation} \hbox{ \strut on intentional programming languages.}}                                                     \\ \hline

\vtop{\hbox{ \strut Huda } \hbox{ \strut Mohamed}}        & DMARF & \vtop{\hbox{ \strut {[}1{]}Autonomic specification of} \hbox{ \strut self-protection for Distributed} \hbox{ \strut MARF with ASSL}}                                                        \\ \cline{2-3} 
                    & GIPSY & \vtop{\hbox{ \strut {[}7{]}Unifying and refactoring DMF} \hbox{ \strut to support concurrent Jini and} \hbox{ \strut JMS DMS in GIPSY. }}                                                   \\ \hline

\vtop{\hbox{ \strut Jilson } \hbox{ \strut Thomas}}       & DMARF & \vtop{\hbox{ \strut {[}6{]}Towards security hardening of scientific} \hbox{ \strut distributed demand-driven and } \hbox{ \strut pipelined computing systems}}                               \\ \cline{2-3} 

                    & GIPSY &  \vtop{\hbox{ \strut {[}10{]}Advances in the design and} \hbox{ \strut implementation of a multi-tier architecture } \hbox{ \strut in the GIPSY environment with Java. }}                     \\ \hline

\vtop{\hbox{ \strut Zakaria} \hbox{ \strut Alomari} }   & DMARF & \vtop{\hbox{ \strut {[}4{]}Distributed Modular Audio Recognition } \hbox{ \strut Framework (DMARF) and its applications } \hbox{ \strut over web services. }}                                 \\ \cline{2-3} 
                    & GIPSY & \vtop{\hbox{ \strut {[}12{]}Towards a self-forensics property} \hbox{ \strut in the ASSL toolset. }}                                                                         \\ \hline
\end{tabular}
\end{table}

%\section*{Design Pattern Distribution}

 \begin{table}[htpb]
    
  \renewcommand{\arraystretch}{1.3}
 \caption{Design Pattern Distribution: Third Deliverable} 

    \begin{tabular}{|c|c|}
%{|p{3cm} | p{2cm} | p{2cm}|}

    \hline
    Name & Design Pattern \\ \hline
     Abdulrhman Albeladi &  Factory   \\ \hline

  Aber Abozkhar  &  Singleton
 \\ \hline
 
  Ahmed Almessabi &  Observer
 \\ \hline

 Huda Mohamed &  Strategy
 \\ \hline

 Jilson Thomas &  Facade
 \\ \hline

 Zakaria Alomari&  Adapter
 \\ \hline

    \end{tabular}

\label{tab:2}

 \end{table}

%\section*{Terminology}
\begin{table}[htpb]
    
  \renewcommand{\arraystretch}{1.3}
 \caption{Terminology} 

    \begin{tabular}{|c|p{7cm}|}
 \hline
Term & Definition                                     \\\hline
MARF     & Modular Audio Recognition Framework                                     \\\hline
DMARF    & Distributed MARF                                                        \\\hline
NLP      & Natural Language Processing                                             \\\hline
CORBA    & Common Object Request Broker Architecture                               \\\hline
XML-RPC  & Remote Procedure Call (RPC) protocol which uses XML to encode its calls \\\hline
Java RMI & Java Remote Method Invocation                                           \\\hline
API      & Application Programmers Interface                                       \\\hline
PDAs     & personal digital assistant                                              \\\hline
SNMP     & Simple Network Management Protocol                                      \\\hline
TCP      & Transmission Control Protocol                                           \\\hline
RMI      & Remote Method Invocation                                                \\\hline
WS       & Web Services                                                            \\\hline
ASSL     & Autonomic System Specification Language                                 \\\hline
GIPSY    & General Intensional Programming System                                  \\\hline
AS       & Autonomic System                                                        \\\hline
AE       & Autonomic Elements                                                      \\\hline
ASIP     & AS Interaction Protocol                                                 \\\hline
GIPC     & General Intensional Programming Compiler                                \\\hline
GEE      & General Eduction Engine                                                 \\\hline
JVM      & The Java Virtual Machine.                                               \\\hline
RPC      & Remote Procedure Call                                                  \\\hline
WSDL     & Web Services Definition Language                                      
 \\\hline
\end{tabular}
\end{table}

\end{document}